\documentclass[reprint,amsmath,amssymb,pra,aps,twocolumn,showpacs]{revtex4-1}

\usepackage{mathrsfs}
\usepackage{graphicx}
\usepackage{subfigure}
\usepackage{amsmath}
\usepackage{amssymb}
\usepackage{amsfonts}
\usepackage{color}
\usepackage{leftidx}
\usepackage{tikz}

\begin{document}
	
\title{Efficient Generation of Spin Cat States}

\author{Jiahao Huang}
\altaffiliation{Email: hjiahao@mail2.sysu.edu.cn, eqjiahao@gmail.com}
\affiliation{Guangdong Provincial Key Laboratory of Quantum Metrology and Sensing $\&$ School of Physics and Astronomy, Sun Yat-Sen University (Zhuhai Campus), Zhuhai 519082, China}
\affiliation{State Key Laboratory of Optoelectronic Materials and Technologies, Sun Yat-Sen University (Guangzhou Campus), Guangzhou 510275, China}

\author{Hongtao Huo}
\affiliation{Guangdong Provincial Key Laboratory of Quantum Metrology and Sensing $\&$ School of Physics and Astronomy, Sun Yat-Sen University (Zhuhai Campus), Zhuhai 519082, China}
\affiliation{State Key Laboratory of Optoelectronic Materials and Technologies, Sun Yat-Sen University (Guangzhou Campus), Guangzhou 510275, China}

\author{Min Zhuang}
\affiliation{Guangdong Provincial Key Laboratory of Quantum Metrology and Sensing $\&$ School of Physics and Astronomy, Sun Yat-Sen University (Zhuhai Campus), Zhuhai 519082, China}
\affiliation{State Key Laboratory of Optoelectronic Materials and Technologies, Sun Yat-Sen University (Guangzhou Campus), Guangzhou 510275, China}

\author{Chaohong Lee}
\affiliation{Guangdong Provincial Key Laboratory of Quantum Metrology and Sensing $\&$ School of Physics and Astronomy, Sun Yat-Sen University (Zhuhai Campus), Zhuhai 519082, China}
\affiliation{State Key Laboratory of Optoelectronic Materials and Technologies, Sun Yat-Sen University (Guangzhou Campus), Guangzhou 510275, China}
	
\begin{abstract}
Spin cat states are promising candidates for achieving Heisenberg-limited quantum metrology.
It is suggested that spin cat states can be generated by adiabatic evolution.
However, due to the limited coherence time, the adiabatic process may be too slow to be practical.
To speed up the state generation, we propose to use machine optimization to generate desired spin cat states.
Our proposed scheme relies only on experimentally demonstrated one-axis twisting interactions with piecewise time-modulation of rotations designed via machine optimization.
The required evolution time is much shorter than the one with adiabatic evolution and it does not make large modification to the existing experimental setups.
Our protocol with machine optimization is efficient and easy to be implemented in state-of-the-art experiments.	
\end{abstract}	

\date{\today}

\maketitle

\section{Introduction}\label{Sec1}
Quantum metrology can achieve higher measurement precision than classical metrology~\cite{Pezze2018,Degen2017} and it plays a central role in quantum science and technology~\cite{Szigeti2021}.
Quantum metrology exploits quantum effects such as entanglement and quantum correlation to achieve better measurement precision~\cite{Giovannetti2006, Lee2006, Pezze2009}.
Many proposals and proof-of-principle demonstrations suggest that entanglement can be used for improving the measurement precision of quantum sensors~\cite{Ockeloen2013,Hosten2016,Szigeti2020}.
For an ensemble of $N$ uncorrelated atoms, the measurement precision can only reach the standard quantum limit (SQL)~\cite{Giovannetti2004}, which is the basic statistical scaling of $1/\sqrt{N}$.
By employing atom-atom entanglement, entangled states such as spin squeezed state~\cite{Kitagawa1993, Wineland1992, Wineland1994, Ma2011, Sackett2010}, Greenberger-Horne-Zeilinger (GHZ) and NOON states~\cite{Lee2006, Pezze2009, Monz2011}, twin-Fock states~\cite{Dunningham2002,Campos2003,Gerry2010, Luo2017} and spin cat states~\cite{Huang2015, Lu2019} can beat the SQL or even approach the fundamental limit of quantum metrology, the Heisenberg limit (HL)~\cite{Huang2014,Degen2017} with a scaling of $1/N$.

Spin cat states are promising candidates for approaching the HL~\cite{Huang2018-1}.
It has been shown that spin cat states with modest entanglement can perform high-precision phase measurement beyond the SQL even under dissipation~\cite{Huang2015}.
Combined with interaction-based readout~\cite{Nolan2017}, spin cat states can perform Heisenberg-limited phase estimation only via population measurement and the robustness against detection noise can be better than using spin squeezed states~\cite{Huang2018-1}.
The high measurement precision and good robustness makes spin cat states appealing for quantum metrology.

The main challenge against the applications of spin cat states in practice is their generation in realistic experiments.
An efficient way to generate spin cat states is the adiabatic evolution~\cite{Lee2006, Lee2009, Gross2012-1, Gross2012-2, Huang2015, Huang2018-2, Zhuang2020}.
By adiabatic sweeping the control parameter across the spontaneous symmetry breaking transition~\cite{Lee2006, Lee2009, Trenkwalder2016}, spin cat states with different degree of entanglement can be prepared~\cite{Huang2015, Huang2018-2}.
However, the adiabatic sweeping demands the control parameter varies slowly to let the evolved state follow the instant ground state of the system.
Due to the limited coherence time in experiments, adiabatic process is too time-consuming and hard to realize.
Hence, developing efficient approach for creating spin cat states becomes a significant issue.

One-axis twisting (OAT) is one of the well-known strategies for generating entanglement in many-body quantum systems.
Under the time-evolution of OAT interaction, an initial spin coherent state (without entanglement) can evolve to spin squeezed states~\cite{Kitagawa1993,Jin2009}, over-squeezed states~\cite{Strobel2014,Davis2016} and other kinds of nonclassical states~\cite{Molmer1999,Micheli2003,Pezze2018}.
%
%
To accelerate state preparation process, twist-and-turn (TNT) dynamics is proposed in which an additional rotation is introduced along with OAT interaction~\cite{Muessel2015,Mirkhalaf2018,Sorelli2019}.
Faster entanglement generation (e.g., spin squeezing) has already been demonstrated experimentally~\cite{Muessel2015}.
Recently, a protocol based on TNT dynamics with a machine-designed time-dependent rotation sequence is proposed~\cite{Haine2020}, in which a higher sensitivity for concurrent entanglement generation and sensing can be achieved~\cite{Haine2020,Huo2022}.
One may also use TNT dynamics with time-dependent rotation sequence for preparation of spin cat states.
It is natural to treat the state preparation process as an optimal control problem~\cite{Omran2019,Liu2021,Lin2021, Kudra2021,Carrasco2022}.
Can one use machine optimization to design suitable time-sequence of control parameter to speed up the generation of spin cat states?

In this article, we investigate how to efficiently generate spin cat states.
At first, we illustrate how to prepare spin cat states via adiabatic evolution.
We show that the recent proposal of adiabatic-parameter-fixed sweeping~\cite{Zhuang2020} can generate spin cat states with high fidelity.
However the adiabatic process may be still too slow to be practical.
To speed up the state generation, we propose to use machine optimization to generate a desired spin cat state based on TNT dynamics.
Our proposed scheme relies only on the developed OAT interactions with piecewise time-modulation of rotations designed via machine optimization.
Compared with adiabatic evolution, the fidelity to the spin cat states can be higher along with the required evolution time becomes much shorter.
It does not require large modification to the existing experimental setups.
Our scheme can be realized with state-of-the-art techniques in a Bose-Einstein condensate (BEC) system~\cite{Riedel2010,Gross2010} or an optical cavity system with light-mediated interactions~\cite{Davis2016,Colombo2021,Li2021}.
It points out an alternative way for generating various entangled states, which has a broad interest for quantum technologies such as quantum metrology and quantum computing.

\section{Model and spin cat states}\label{Sec2}
\subsection{Ensemble of Bose atoms}
An ensemble of $N$ two-level atoms can be regraded as $N$ identical spin-$\frac {1}{2}$ particles.
The system can be conveniently described by a collective spin with spin length $J=\frac {N}{2}$ and the states can be displayed on a generalized Bloch sphere.
The collective spin operator $\hat{J}$ contains three components,  $\hat{J_x}=\frac {1}{2}(\hat{a}^{\dag}\hat{b}+\hat{b}^{\dag}\hat{a})$, $\hat{J_y}=\frac {i}{2}(\hat{a}^{\dag}\hat{b}-\hat{b}^{\dag}\hat{a})$, $\hat{J_z}=\frac {1}{2}(\hat{b}^{\dag}\hat{b}-\hat{a}^{\dag}\hat{a})$ , where $\hat{a}$ and $\hat{b}$ are the annihilation operators for particles in level $|a\rangle$ and $|b\rangle$, respectively.
These collective spin operators obey the general angular momentum commutation relations $[\hat{J_i},\hat{J_j}]=i\hbar\epsilon_{ijk}\hat{J_k}$ with $i,j,k=x,y,z$ and $\epsilon_{ijk}$ the Levi-Civita symbol.
In this representation, a system state can be expressed by $|\Psi\rangle=\sum_{m=-N/2}^{N/2} C_{m}|J, m\rangle$, where $|J, m\rangle$ is the Dicke basis denoting $N/2-m$ particles in $|a\rangle$ and $N/2+m$ particles in $|b\rangle$.

\subsection{Spin cat states}
Spin cat states are excellent candidates for achieving Heisenberg-limited phase estimation.
Spin cat state is a typical kind of macroscopic superposition of spin coherent states (MSSCS).
Generally, an MSSCS is a superposition of multiple spin coherent states (SCSs)~\cite{Micheli2003, Ferrini2010, Spehner2014}, which can be written in the form of $|\Psi(\theta, \varphi)\rangle_{\textrm{M}}=\mathcal{N}_{C}(|\theta,\varphi\rangle + |\pi-\theta,\varphi\rangle)$,
where $\mathcal{N}_{C}$ is the normalization factor and $\left|\theta,\varphi\right\rangle$ denotes the spin coherent state (SCS) with $\left|\theta,\varphi\right\rangle=\sum_{m}\! \sqrt{\frac{(2J)!}{(J+m)!(J-m)!}}\! \cos^{J+m}\!\left({\theta \over 2}\right)\!\sin^{J-m}\!\left({\theta \over 2}\right)\! e^{-i(J+m)\varphi} \left|J,m\right\rangle$.
Since $c_m(\theta)=c_{-m}(\pi-\theta)$, the coefficients are symmetric about $m=0$. We assume the two SCSs have the same azimuthal angle $\varphi=0$, $|\Psi(\theta)\rangle_{\textrm{M}}=\mathcal{N}_{C} \left[\sum^{J}_{m=-J} c_m(\theta)\left(\left|J,m\right\rangle+\left|J,-m\right\rangle\right)\right]$.

The properties of the MSSCS depends on $\theta$.
When $\theta=\pi/2$, it corresponds to a SCS $\left|\pi/2,0\right\rangle$.
As $\theta$ decreases, the two superposition SCSs become separated. When $\theta \lesssim \theta_c\equiv\sin^{-1}\left\{2\left[\frac{\left((J-1)!\right)^2}{2 (2J)!}\right]^{1\over{2J}}\right\}$ is sufficiently small~\cite{Huang2015}, the two SCSs become quasi-orthogonal (or orthogonal), the MSSCS can be regarded as a spin cat state.  %
In this case, we abbreviate the spin cat states as $\left|\Psi(\theta)\right\rangle_{\textrm{CAT}}$, and it can be approximated as
\begin{eqnarray}\label{CAT}
    \left|\Psi(\theta)\right\rangle_{\textrm{CAT}}&\approx& \frac{1}{\sqrt{2}}\left(|\theta,\varphi\rangle + |\pi-\theta,\varphi\rangle\right)\\\nonumber
    &=&\frac{1}{\sqrt{2}} \left[\sum^{J}_{m=-J} c_m(\theta)\left(\left|J,m\right\rangle+\left|J,-m\right\rangle\right)\right].
\end{eqnarray}
Note that spin cat states can be understood as a superposition of GHZ states with different spin length.
Particularly when $\theta=0$, it reduces to a GHZ state, i.e., $\left|\Psi(0)\right\rangle_{\textrm{CAT}}=\frac{1}{\sqrt 2}\left(|J,-J\rangle + |J,J\rangle \right)$.
The expectation of $\hat J_z$ for a spin cat state $_\textrm{CAT}\langle\Psi(\theta) |\hat J_z| \Psi(\theta)\rangle_\textrm{CAT}=\sum_{m} m c_m^{*}(\theta) c_m(\theta)=0$.
Hence, the variance of a spin cat state becomes $\Delta^2 \hat J_z=\langle\Psi(\theta) |\hat J_z^2| \Psi(\theta)\rangle_\textrm{CAT} = \sum_{m} m^2 c_m^{*}(\theta) c_m(\theta)$.
The variance can be analytically obtained $\Delta^2 \hat J_z \approx \frac{1}{4} N^2 \cos^2\theta$, which only depends on $\theta$ and $N$.

If a spin cat state $\left|\Psi(\theta)\right\rangle_{\textrm{CAT}}$ is input for interferometry, and it undergoes a unitary evolution $|\psi_{out}\rangle = U(\phi)\left|\Psi(\theta)\right\rangle_{\textrm{CAT}}=e^{-i\hat J_z \phi} \left|\Psi(\theta)\right\rangle_{\textrm{CAT}}$, the quantum Fisher information (QFI) for the output state~\cite{Braunstein1994, Huang2014} can be calculated as $F^Q_\textrm{CAT}=4\left( \langle \psi'|\psi'\rangle - |\langle \psi'|\psi_{out}\rangle|^2\right)=4\Delta^2 \hat{J}_z \approx {N^2}\cos^2 \theta$ with $|\psi'\rangle=d |\psi_{out}\rangle/d\phi$.
Thus the ultimate phase precision by a spin cat state is obtained,
\begin{eqnarray}\label{QCRB_M}
    \Delta\phi \ge \Delta\phi_{Q}\equiv\frac{1}{\sqrt{F^Q_\textrm{CAT}}} = \frac{1} {N \cos \theta}.
\end{eqnarray}

According to the ultimate measurement bound~\eqref{QCRB_M}, the achievable precision of the spin cat state $|\Psi(\theta)\rangle_\textrm{CAT}$ follows the Heisenberg scaling multiplied by a coefficient $1/\cos\theta$ only dependent on $\theta$.
When $\theta=0$, $|\Psi(0)\rangle_\textrm{CAT}$ is the GHZ state, its ultimate bound is the exact Heisenberg limit $1/N$.
When $0<\theta\lesssim \theta_c$, the ultimate bound becomes $1/(N\cos\theta)$, still has the Heisenberg limited scaling.
For example, $\Delta\phi_{Q}=2/\sqrt{3}N, \sqrt{2}/N, 2/N$ for spin cat states $|\Psi(\theta)\rangle_\textrm{CAT}$ with $\theta=\pi/6, \pi/4, \pi/3$.

\subsection{State generation}
For entangled state preparation in an atomic ensemble, one can use the Bose-Josephson Hamiltonian~\cite{Strobel2014} (we set $\hbar=1$ hereafter),
\begin{equation}\label{Ham}
	\hat{H}(t)=\chi\hat{J_z^2}+\Omega(t) \hat{J_x}.
\end{equation}
Here, $\chi$ denotes the magnitude of the twisting strength and $\Omega(t)$ is Rabi frequency determining the rotation rate around the $\hat J_{x}$ axis.
The first term $\chi\hat{J_z^2}$ is the consequence of twisting dynamics and creates the entanglement among atoms.
The second term $\Omega(t) \hat{J_x}$ rotates the collective spin state in a perpendicular direction.
In the following calculations, we set the parameters in the unit of $|\chi|$.

For a state preparation process with an initial pure state $|\psi_{0}\rangle$, the final prepared state at time $T$ can be expressed as $\psi(T)=\int_{0}^{T} e^{-i \hat H(t) dt} |\psi_{0}\rangle$.
To characterize how close between the prepared state $\psi(T)$ and a desired state $|\psi_{d}\rangle$, one can introduce the fidelity, which is expressed as
\begin{equation}
	 F(T)= |\langle \psi(T)|\psi_{d}\rangle|^2.
\end{equation}
When $|\psi(T)\rangle=|\psi_{d}\rangle$, $F(T)=1$. While $F(T)=0$ if $|\psi(T)\rangle$ is orthogonal with $|\psi_{d}\rangle$.
Generally, one can start from an initial SCS $|\psi_{0}\rangle=\left|\pi/2,0\right\rangle$, which is a non-entangled state and easy to prepare in experiments.
In the following, we show how to generate a desired spin cat state $|\psi_{d}\rangle=|\Psi(\theta)\rangle_{\rm{CAT}}$ with $F(T)=|\langle \psi(T)|\Psi(\theta)\rangle_{\rm{CAT}}|^2$ close to $1$ by designing time-dependent modulation of $\Omega(t)$.

\section{spin cat state generation via adiabatic evolution}
\begin{figure}[htb]
\centering
\includegraphics[width=\columnwidth]{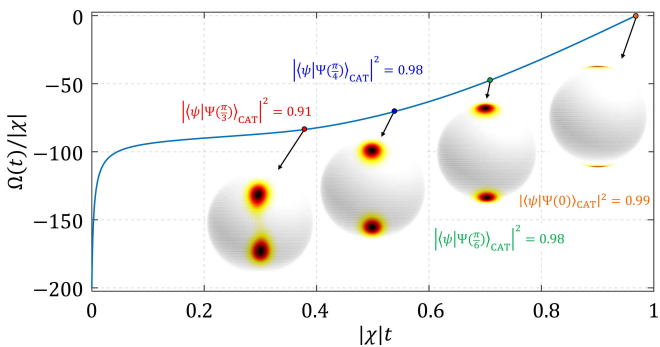}\caption{(Color online) Spin cat state generatin via adibatic evolution with adiabatic-parameter-fixed sweeping. Blue solid line is the variation of Rabi frequency $\Omega(t)$ versus evolution time $t$. Here, the total atom number $N=100$ and $\chi<0$. For $|\chi|t = 0.375, 0.55, 0.69, 0.96$, the corresponding evolved states are close to the spin cat states of $|\Psi(0)\rangle_{\rm{CAT}}$, $|\Psi(\frac{\pi}{6})\rangle_{\rm{CAT}}$, $|\Psi(\frac{\pi}{4})\rangle_{\rm{CAT}}$, $|\Psi(\frac{\pi}{3})\rangle_{\rm{CAT}}$ with fidelity $F=0.99, 0.98, 0.98, 0.91$, respectively.}
\label{Fig1}
\end{figure}

For the Hamiltonian~\eqref{Ham}, when $|\Omega/\chi| \gg 1$, the system ground state is an SU(2) SCS.
The sign of $\chi$ determines the properties of the ground state when $\Omega$ is not large enough.
When $\chi>0$ and $|\Omega/\chi| \ll 1$, the ground state is a spin squeezed state.
While for $\chi<0$, the ground state becomes a spin cat state when $|\Omega/\chi| \ll 1$.
To generate a spin cat state, it is natural to prepare the ground state of Hamiltonian~\eqref{Ham} in the limit of $|\Omega/\chi| \ll 1$ with $\chi<0$.
For atomic BEC system, the negative twisting strength can be achieved by tuning the interspecies s-wave scattering length via Feshbach resonance~\cite{Gross2012-1, Gross2012-2}.

\begin{figure}[htb]
\centering
\includegraphics[width=\columnwidth]{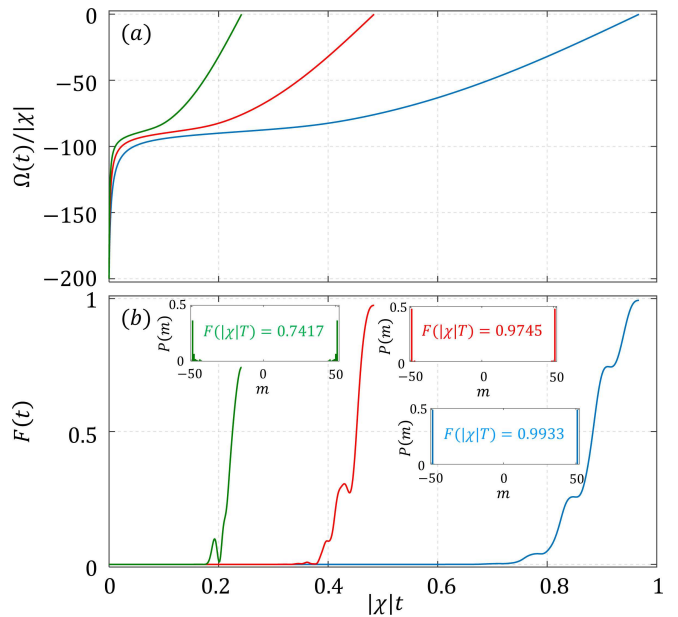}\caption{(Color online) Spin cat state generation via adiabatic-
parameter-fixed sweeping with different $\epsilon$. Green, red and blue lines represent the results of $\epsilon=0.2, 0.1,$ and $0.05$, respectively. (a) The variation of Rabi frequency $\Omega(t)$ versus evolution time $t$. (b) The fidelity $F(t)=|\langle \psi(t)|\Psi(0)\rangle_{\rm{CAT}}|^2$ between the instant evolved state and the target spin cat state $|\Psi(0)\rangle_{\rm{CAT}}$ versus time $t$. The insets show the probability distribution $P(m)=|\langle J,m|\psi(T)\rangle|^2$ of the final prepared states. Here, the total atom number $N=100$ and $\chi<0$.}
\label{Fig2}
\end{figure}

This model is invariant under the transformation of exchanging mode $a$ and mode $b$~\cite{Zhuang2020}.
Under the transformation $\hat a (\hat b) \rightarrow \hat b (\hat a)$, $\hat J_x \rightarrow \hat J_x$, $\hat J_z \rightarrow -\hat J_z$, hence the Hamiltonian~\eqref{Ham} remains unchanged.
Thus, this system possesses a parity symmetry and it guarantees the symmetry-protected adiabatic evolution~\cite{Zhuang2020}.
The adiabatic evolution can be happened since there is always a finite minimum energy gap between instantaneous eigenstates of the same parity.
With negative $\chi$, one possibility for generating spin cat states is the adiabatic evolution.
Initially, we set $\Omega(0)$ to be sufficiently large, the ground state is nearly an SCS along $x$ axis with even parity.
By sweeping the Rabi frequency across the critical point ${\Omega_c}/{N|\chi|}=1$, the two lowest eigenstates change from non-degenerate to degenerate.
Through adiabatically sweeping $\Omega(t)/\chi$ to zero, the evolved state will stay in the instant ground state and spin cat states (also with even parity) can be prepared when $\Omega(T)/\chi$ close to 0.

Naively, one can linearly sweep $\Omega(t)=\Omega(0)+\upsilon t$ from the non-degenerate regime across to the degenerate regime with fixed the sweeping rate $\upsilon$. If $\upsilon$ is sufficiently small, the adiabatic evolution of the ground state can still be achieved with high fidelity.
However, this linear sweeping scheme is not timesaving.

\begin{figure*}[htb]
\centering
\includegraphics[width=2\columnwidth]{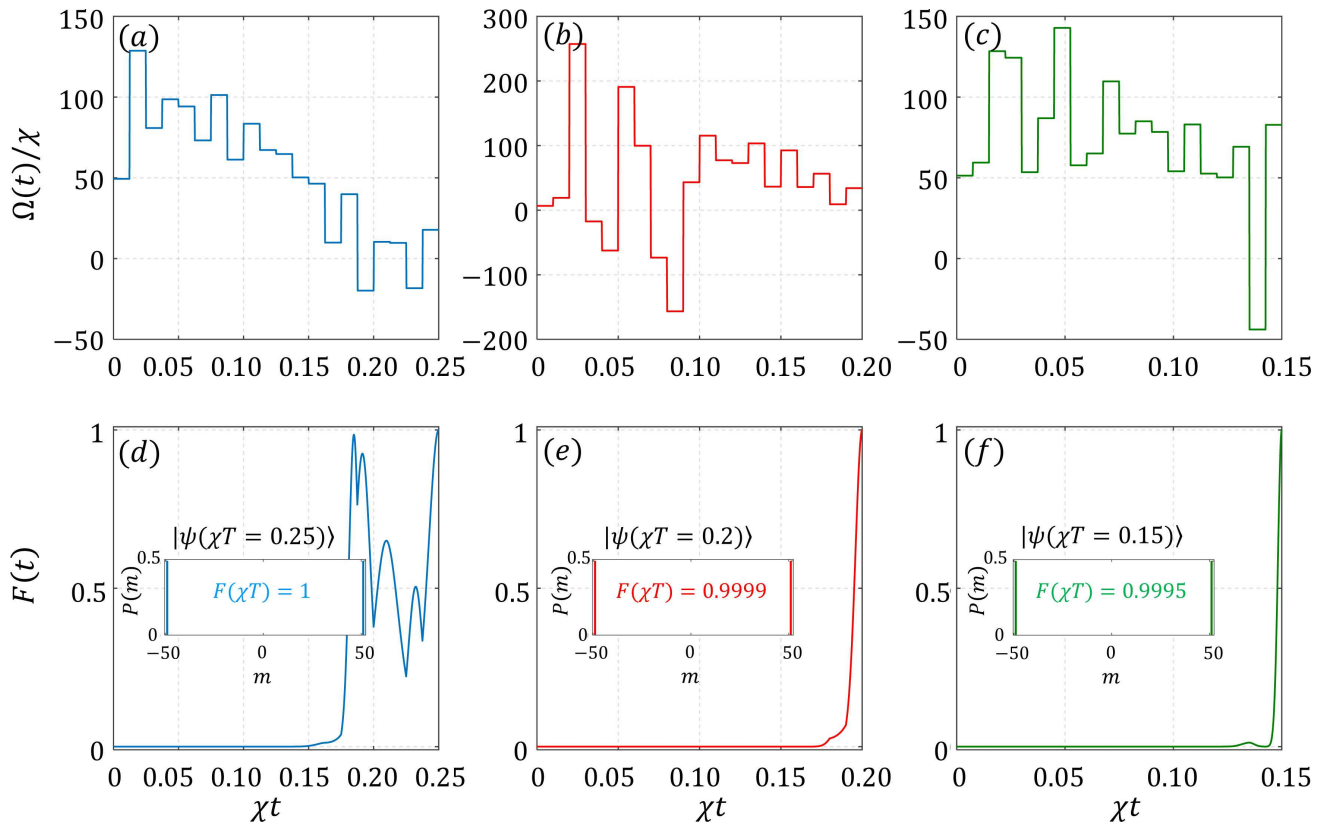}\caption{(Color online) Spin cat state generation via machine optimization. The optimized time-sequence of Rabi frequency $\Omega(t)$ for total evolution time (a) $\chi T=0.25$, (b) $\chi T=0.2$, and (c) $\chi T=0.15$. The corresponding fidelity $F(t)=|\langle \psi(t)|\Psi(0)\rangle_{\rm{CAT}}|^2$ between the instant evolved state and the target spin cat state $|\Psi(0)\rangle_{\rm{CAT}}$ versus time $t$ are shown in (d), (e) and (f). The insets show the probability distribution $P(m)=|\langle J,m|\psi(T)\rangle|^2$ of the final prepared states. Here, the segment number is chosen as $n=20$, the total atom number $N=100$ and $\chi>0$.}
\label{Fig3}
\end{figure*}

To perform faster ground state adiabatic evolution, we change the sweeping rate with time according to the instantaneous energy gaps between the ground state and the second excited state (both with the same parity) under a fixed adiabatic parameter $\epsilon$.
%
Since $\epsilon$ is fixed, we call it adiabatic-parameter-fixed sweeping~\cite{Zhuang2020}.
The time-varying Rabi frequency $\Omega(t)=\Omega(0)+\int_{0}^{t} \upsilon(t') dt'$, where $\upsilon(t)=\dot{\Omega}(t)$ is the instant sweeping rate of the Rabi frequency.
For adiabatic-parameter-fixed sweeping, $\upsilon(t) = \frac{\epsilon{\left[{E_{1}}(t)-E_{3}(t)\right]}^{2}} {\left|\langle{\phi_{1}(t)}|\hat J_x| {\phi_{3}(t)}\rangle\right|}$. Here, $E_1(t)$ and $E_3$ respectively represent the energy of instant ground state $\phi_{1}(t)$ and the second excited state $\phi_{3}(t)$ of Hamiltonian~\eqref{Ham}.
The modulation of Rabi frequency $\Omega(t)$ with time is shown in Fig.~\ref{Fig1}.
Based on the adiabatic-parameter-fixed sweeping scheme, the total time for adiabatic evolution can be reduced compared with the naive linear sweeping~\cite{Zhuang2020}.
Besides, the generated states and their fidelities to the target spin cat states are also shown in Fig.~\ref{Fig1}.
The fidelities can be above 0.9 under the condition of $\epsilon=0.05$ with total atom number $N=100$.

The adiabatic parameter $\epsilon$ determines the efficiency of the sweeping. With smaller $\epsilon$, the fidelity to the target spin cat state can be higher, however the total evolution time $T$ will be longer.
We choose $\epsilon=0.05, 0.1, 0.2$ for illustration.
If the target spin cat state is the GHZ state $|\Psi(0)\rangle_{\rm{CAT}}$, the best fidelity can be $0.74, 0.97, 0.99$ with $\epsilon=0.2, 0.1, 0.05$, respectively.
The corresponding total evolution time are $\chi T=0.24, 0.48, 0.96$.
That is, to generate a GHZ state ($N=100$) with fidelity over 0.99, $\chi T>0.9$ should be necessary.

\section{spin cat state generation via machine optimization for quantum metrology}
Although spin cat states can be generated by adiabatic quantum evolution, however the adiabatic process is always too slow to be practical.
In this section, we demonstrate how to use machine optimization to generate a desired spin cat state.
Here, we propose a different method based on TNT dynamics.
The presence of a nonzero constant term $\Omega \hat J_x$ can generate entanglement more rapidly than OAT.
Our proposed scheme makes use of the same operations as the TNT implementations, but with a time-dependent Rabi frequency sequence $\Omega(t)$ designed via machine optimization.
The required evolution time is much shorter than the one with adiabatic evolution.
Finally, we also use the prepared states for phase estimation and the ultimate precision bounds follow the Heisenberg-limited scalings, which are consistent with the analytical analysis.

\begin{figure*}[htb]
\centering
\includegraphics[width=2\columnwidth]{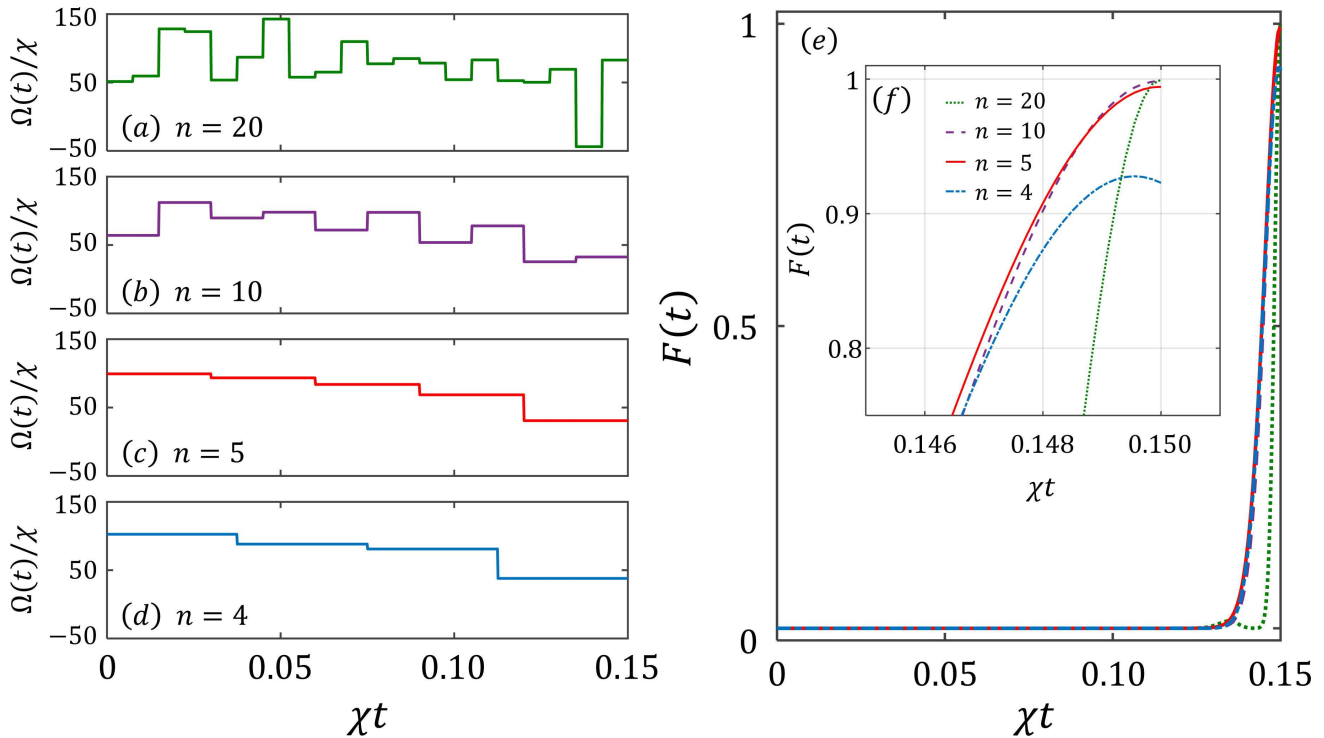}\caption{(Color online) Spin cat state generation via machine optimization for a fixed total evolution time $T$ with different segment number $n$. Here, the total evolution time $\chi T=0.15$, the total atom number $N=100$ and $\chi>0$. The optimized time-sequence of Rabi frequency $\Omega(t)$ for segment number (a) $n=20$, (b) $n=10$, (c) $n=5$, and (d) $n=4$. The corresponding fidelity $F(t)=|\langle \psi(t)|\Psi(0)\rangle_{\rm{CAT}}|^2$ between the instant evolved state and the target spin cat state $|\Psi(0)\rangle_{\rm{CAT}}$ versus time $t$ are shown in (e). Inset is the enlarged region near the final time $T$.}
\label{Fig4}
\end{figure*}

\subsection{State generation via machine optimization}
We still start from an initial SCS and design the time-sequence of Rabi frequency via the techniques of optimization. Without loss of generality, we consider the twisting strength $\chi$ to be positive, which is commonly used for entangled state generation such as spin squeezed states.
For the Hamiltonian~\eqref{Ham} with total evolution time $T$, we consider $T$ to be divided into $n$ equal segments so that the time-dependent Rabi frequency can be parameterized as~\cite{Haine2020}
\begin{equation}\label{Omegax}
	\Omega(t)=\Lambda(t)N\chi/2
\end{equation}
with $\Lambda(t)$ a piecewise step function and in each segment ($k=1,2,...,n$)
\begin{equation}\label{Lambdax}
	\Lambda(t)=\Lambda^{(k)}, \quad (k-1)T/n\le t <kT/n
\end{equation}
can be varied individually.
Thus, the time-dependent variable $\Omega(t)$ involves $n$ variational parameters. It is assumed that these parameters can be varied arbitrarily.
Given $\chi T$ and $N$, starting from an initial SCS $|\frac{\pi}{2},0\rangle$ one can obtain the maximal fidelity $F(T)$ by optimizing $\Omega(t)$.
Here, we use the minimization routine in Matlab with the BFGS method to iteratively search the parameters $\Lambda^{(k)}$ that minimizing $-F(T)$.

We first consider the segment number $n=20$ and the target spin cat state is a GHZ state $|\Psi(0)\rangle_{\rm{CAT}}$.
We show the results with $\chi T=0.25, 0.20, 0.15$ in Fig.~\ref{Fig3}.
As shown, the final prepared states can all arrive to the desired GHZ state with fidelity over $0.999$.
The optimized sequences of $\Omega(t)$ are shown in the top row, and the corresponding evolutions of fidelity are depicted in the bottom row.
Note that, the final fidelity $F(\chi T)=1, 0.9999, 0.9995$ for $\chi T=0.25, 0.20, 0.15$, respectively.
The longer total evolution time, the larger final fidelity it can attain, which is consistent with our common intuition.

More importantly, compared with the scheme of adiabatic evolution, the scheme via machine optimization requires much shorter evolution time $T$.
To achieve the GHZ state with fidelity over $0.99$, $|\chi|T$ should be larger than $0.9$ for adiabatic-parameter-fixed sweeping.
However, for machine optimization, $\chi T$ can be as small as $0.15$, which saves up to $80\%$ time for state generation.
In addition, the variation of $\Omega(t)$ is no longer a continuous time-dependent modulation.
Instead, it becomes a step-wise variation with several segments, which is easier to realize in experiments.
In practice, the fewer segment number it requires, the more convenient the experiment will be.

\begin{figure*}[htb]
\centering
\includegraphics[width=2\columnwidth]{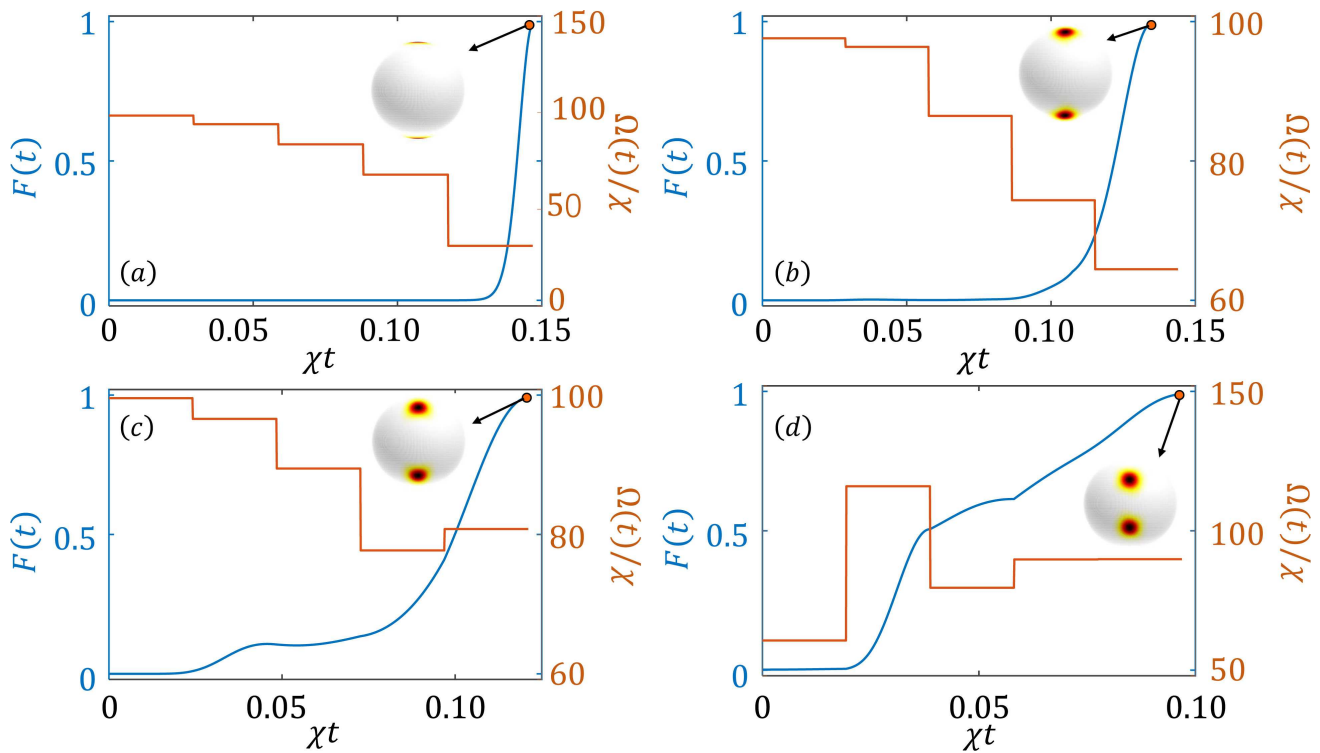}\caption{(Color online) State generation via machine optimization for target spin cat state $|\Psi(\theta)\rangle_{\rm{CAT}}$ with (a) $\theta=0$, (b) $\theta=\frac{\pi}{6}$, (c) $\theta=\frac{\pi}{4}$ and (d) $\theta=\frac{\pi}{3}$. The orange lines are the optimized time-sequence of Rabi frequency $\Omega(t)$. The blue lines are the corresponding fidelity $F(t)=|\langle \psi(t)|\Psi(\theta)\rangle_{\rm{CAT}}|^2$ between the instant evolved state and the target state versus time $t$. Here, the total atom number $N=100$ and $\chi>0$. The total evolution time for (a) $T=0.147$, (b) $T=0.134$, (c) $T=0.121$, and (d) $T=0.097$.}
\label{Fig5}
\end{figure*}

For a fixed total evolution time $\chi T$, there may be a suitable segment number $n$ for optimization.
Generally, if segment number is too small, the degree of freedom may not be sufficient to drive the SCS into the desired state.
While for large segment number, it adds the difficulty for the numerical search since the variational parameters become too many (especially for large atom number).
Besides, it is also experimentally feasible if the segment number is not large.
Thus, there will be a trade-off for choosing the segment number $n$.
Here, we choose $n=4, 5, 10, 20$ with $N=100$ for illustration, see Fig.~\ref{Fig4}.
The target state is still the GHZ state $|\Psi(0)\rangle_{\rm{CAT}}$.
Within the same $\chi T=0.15$, the optimized control for $\Omega(t)$ with $n=5, 10, 20$ can attain the desired state with fidelity over $0.99$.
While for $n=4$, the final fidelity can only reach $F=0.923$.
The final fidelity $F=0.9943, 0.9985, 0.9995$ for $n=5, 10, 20$, respectively.
Despite the fidelity for $n=20$ is the largest, it requires more sophisticated control for $\Omega(t)$.
While for $n=5$ the final fidelity is also over 0.99 and the time-sequence of $\Omega(t)$ is more easy to implement.
As shown, only $n=5$ segments to vary is sufficient to drive the SCS into the desired state and we consider $n=5$ as a suitable segment number.
Therefore, we set $n=5$ for generating other spin cat states with different $\theta$ in the following.

In Fig.~\ref{Fig5}, we show the results of generating $|\Psi(0)\rangle_{\rm{CAT}}$, $|\Psi(\frac{\pi}{6})\rangle_{\rm{CAT}}$, $|\Psi(\frac{\pi}{4})\rangle_{\rm{CAT}}$, and $|\Psi(\frac{\pi}{3})\rangle_{\rm{CAT}}$ via optimizing $\Omega(t)$ with $n=5$.
The orange lines are the optimized sequences of $\Omega(t)$ and the corresponding evolutions of fidelity are depicted with blue lines.
Here, the total evolution time $T$ is chosen as the minimal evolution time that can attain the fidelity over $0.99$.
For $N=100$, it is shown that the spin cat state $|\Psi(\theta)\rangle_{\rm{CAT}}$ with larger $\theta$ requires shorter evolution time $T$, which is easier to generate in practice.
Compared with the results of adiabatic evolution in Fig.~\ref{Fig1}, the generated states via machine optimization are closer to the target spin cat states $|\Psi(\theta)\rangle_{\rm{CAT}}$, especially for $|\Psi(\theta)\rangle_{\rm{CAT}}$ with larger $\theta$ such as $\theta=\frac{\pi}{4}$ and $\frac{\pi}{3}$.
Besides, the total evolution time $\chi T$ is much shorter than the one $|\chi| T$ via adiabatic evolution.

\begin{figure}[htb]
\centering
\includegraphics[width=\columnwidth]{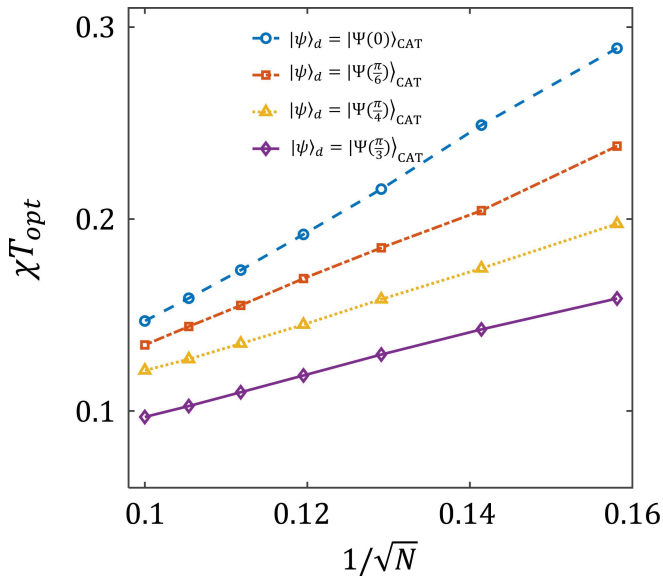}\caption{(Color online) The optimal total evolution time $T_{opt}$ for different spin cat state generation via machine optimization. Here, we fix the segment number $n=5$ and the fidelity between the prepared state and the target state is $F(T_{opt})\gtrsim0.99$. Roughly, the optimal total evolution time $T_{opt}$ exhibits linear relation versus $1/\sqrt{N}$.}
\label{Fig6}
\end{figure}

Fig.~\ref{Fig6} shows the optimal evolution time $T_{opt}$ for generating the spin cat states $|\Psi(\theta)\rangle_{\rm{CAT}}$.
Numerically, we find that the optimal evolution time scales approximately linear versus $1/\sqrt{N}$ for spin cat states.
This relation can guide us to find the total evolution time for generating spin cat states with larger $N$.
For the same twisting strength $\chi$, the required total evolution time $T$ for driving the SCS to a desired spin cat state will become smaller when $N$ gets larger.

\subsection{Ultimate measurement precision}
Finally, we use the prepared states to perform the phase estimation via many-body interferometry.
Conventionally, an input state $|\psi_{in}\rangle$ will evolve into an output state $|\psi_{out}(\phi)\rangle=\hat U (\phi)|\psi_{in}\rangle$ under the unitary transformation $\hat U (\phi)=e^{-i \phi \hat J_z}$ for phase sensing.
In general, for an output state $|\psi_{out}(\phi)\rangle$, the measurement precision is limited by the quantum Cramer-Rao bound (QCRB), i.e.,
\begin{equation}\label{QCRB}
    \Delta\phi \ge \Delta\phi_{Q}\equiv\frac{1}{\sqrt{F^Q}}.
\end{equation}
Here, $F^Q=4\left( \langle \psi'|\psi'\rangle - |\langle \psi'|\psi_{out}(\phi)\rangle|^2\right)$ is the QFI with $|\psi'\rangle=d |\psi_{out}(\phi)\rangle/d\phi$.
Then, we choose the prepared states $|\psi(T)\rangle$ via machine optimization as the input states $|\psi_{in}\rangle$ and calculate the corresponding QCRB $\Delta\phi_{Q}$ according to Eq.~\eqref{QCRB}.

\begin{figure}[htb]
\centering
\includegraphics[width=\columnwidth]{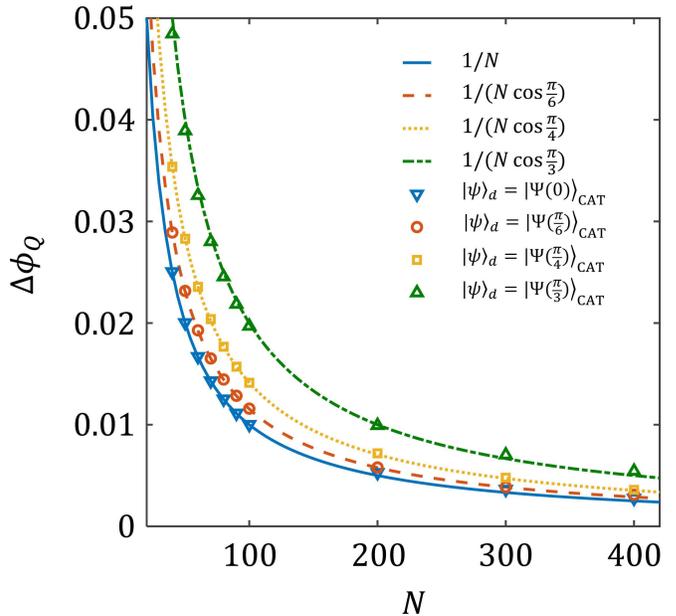}\caption{(Color online) The ultimate measurement precision $\Delta \phi_Q$ with prepared spin cat states versus total atom number $N$. The spin cat states are prepared via machine optimization with segment number $n=5$. The target states for inverted triangles, circles, squares and triangles are $|\Psi(0)\rangle_{\rm{CAT}}$, $|\Psi(\frac{\pi}{6})\rangle_{\rm{CAT}}$, $|\Psi(\frac{\pi}{4})\rangle_{\rm{CAT}}$, $|\Psi(\frac{\pi}{3})\rangle_{\rm{CAT}}$ respectively. The lines represent the corresponding analytical bounds~\eqref{QCRB_M} for spin cat states.}
\label{Fig7}
\end{figure}

The ultimate measurement precisions of the prepared states $\Delta\phi_{Q}$ versus total atom number $N$ are shown in Fig.~\ref{Fig7}.
Since the prepared states have high fidelity with the desired states $|\psi_d\rangle=|\Psi(\theta)\rangle_{\rm{CAT}}$ for $\theta=0, \pi/6, \pi/4, \pi/3$, the ultimate precision bounds are almost the same as the corresponding spin cat states.
For all spin cat states, the ultimate measurement precision is inversely proportional to the total atom number, i.e., $\Delta\phi_{Q} \propto 1/N$ exhibiting the Heisenberg-limited scaling.
As expected, the scaling of spin cat state $|\Psi(\theta)\rangle_{\rm{CAT}}$ versus $N$ is consistent with the analytical bound~\eqref{QCRB_M}.
The measurement precision bound can be saturated by implementing the interaction-based readout~\cite{Huang2018-1}. If another OAT dynamics is performed before the measurement, a Heisenberg-limited measurement precision scaling can be achieved via the population measurement. In addition, compared with the twisting echo schemes with spin squeezed states, the interaction-based readout with spin cat states can be more robust against detection noise~\cite{Huang2018-1}.
This also adds the experimental feasibility to use spin cat states for quantum metrology.


\section{Summary and discussion}\label{Sec5}
In summary, we have shown how to generate spin cat states via twist-and-turn dynamics and machine optimization.
First, spin cat states can be generated via adiabatic evolution.
One can use the adiabatic-parameter-fixed scheme to prepare the spin cat states, which is more efficient than the naive linear sweeping.
However, the adiabatic evolution is still slow to be experimentally practical.
To speed up the process, we have also presented a proposal for generating spin cat states by using the technique of machine optimization.
Instead of continuous time-dependent modulation, only a sequence of step-wise variation with several segments is needed.
More importantly, the required total evolution time is much shorter than the adiabatic evolution.

At last, we briefly discuss the experimental feasibility of our proposal.
The OAT interaction in our scheme had already been realized in experimental platforms of Bose-condensed atoms~\cite{Gross2010} or atomic ensemble interacting with a cavity light field~\cite{Colombo2021}.
For Bose-condensed atoms, the twisting strength which is the effective nonlinearity $\chi \propto g_{aa}+g_{bb}-2g_{ab}$ with $g_{ij}=4\pi\hbar^2 a_{ij}/m$, $a_{ij}$ the s-wave scattering lengths between states $i$ and $j$.
The sign and the amplitude of $\chi$ can be controlled by tuning the interspecies s-wave scattering length $a_{ab}$ via magnetic Feshbach resonance.
For $^{87}$Rb atoms, if the two internal states are selected as $|a\rangle\equiv|F=1,m=1\rangle$ and $|b\rangle\equiv|F=2,m=-1\rangle$, the Feshbach resonance happens at $B=9.092$G~\cite{Gross2012-2}.
When the magnetic field $B=9.13$G the effective nonlinearity $\chi\approx0.4$Hz~\cite{Gross2010}.
For total atom number $N=400$, a spin cat state generated via machine optimization may be achieved within $\chi T \lesssim 0.06$, and the total evolution time can be less than $150$ ms, which is within the BEC lifetime.
For atomic ensemble in cavity, an effective OAT interaction can be produced by utilizing the interaction between atomic ensemble with a single-mode cavity light field.
The sign and the amplitude of $\chi$ can be easily tuned by changing the detuning between the atom-cavity resonance and the light~\cite{Colombo2021}.
While for the rotation operation, the time-modulated controls of Rabi frequency $\Omega(t)$ coupling the two internal states could be implemented by precise controls of the radio-frequency or microwave pulses~\cite{Gross2010}.
Combined with OAT interaction and time modulation of Rabi frequency, our proposal should be easily implemented in experiments.

Our methods can also be extended to generate other significant entangled states such as spin squeezed states~\cite{Carrasco2022} and twin Fock states~\cite{Guo2021}.
It provides a powerful tool and points out a new way for for designing optimal quantum metrology protocols~\cite{Zhou2020,Titum2021,Kaubruegger2021}.

\begin{acknowledgments}
This work is supported by the National Natural Science Foundation of China (12025509, 11874434), the Key-Area Research and Development Program of GuangDong Province (2019B030330001), and the Science and Technology Program of Guangzhou (201904020024). J. H. is partially supported by the Guangzhou Science and Technology Projects (202002030459).
\end{acknowledgments}

%


\begin{thebibliography}{57}%
\makeatletter
\providecommand \@ifxundefined [1]{%
 \@ifx{#1\undefined}
}%
\providecommand \@ifnum [1]{%
 \ifnum #1\expandafter \@firstoftwo
 \else \expandafter \@secondoftwo
 \fi
}%
\providecommand \@ifx [1]{%
 \ifx #1\expandafter \@firstoftwo
 \else \expandafter \@secondoftwo
 \fi
}%
\providecommand \natexlab [1]{#1}%
\providecommand \enquote  [1]{``#1''}%
\providecommand \bibnamefont  [1]{#1}%
\providecommand \bibfnamefont [1]{#1}%
\providecommand \citenamefont [1]{#1}%
\providecommand \href@noop [0]{\@secondoftwo}%
\providecommand \href [0]{\begingroup \@sanitize@url \@href}%
\providecommand \@href[1]{\@@startlink{#1}\@@href}%
\providecommand \@@href[1]{\endgroup#1\@@endlink}%
\providecommand \@sanitize@url [0]{\catcode `\\12\catcode `\$12\catcode
  `\&12\catcode `\#12\catcode `\^12\catcode `\_12\catcode `\%12\relax}%
\providecommand \@@startlink[1]{}%
\providecommand \@@endlink[0]{}%
\providecommand \url  [0]{\begingroup\@sanitize@url \@url }%
\providecommand \@url [1]{\endgroup\@href {#1}{\urlprefix }}%
\providecommand \urlprefix  [0]{URL }%
\providecommand \Eprint [0]{\href }%
\providecommand \doibase [0]{http://dx.doi.org/}%
\providecommand \selectlanguage [0]{\@gobble}%
\providecommand \bibinfo  [0]{\@secondoftwo}%
\providecommand \bibfield  [0]{\@secondoftwo}%
\providecommand \translation [1]{[#1]}%
\providecommand \BibitemOpen [0]{}%
\providecommand \bibitemStop [0]{}%
\providecommand \bibitemNoStop [0]{.\EOS\space}%
\providecommand \EOS [0]{\spacefactor3000\relax}%
\providecommand \BibitemShut  [1]{\csname bibitem#1\endcsname}%
\let\auto@bib@innerbib\@empty
\bibitem [{\citenamefont {Pezz{\`e}}\ \emph {et~al.}(2018)\citenamefont
  {Pezz{\`e}}, \citenamefont {Smerzi}, \citenamefont {Oberthaler},
  \citenamefont {Schmied},\ and\ \citenamefont {Treutlein}}]{Pezze2018}%
  \BibitemOpen
  \bibfield  {author} {\bibinfo {author} {\bibfnamefont {L.}~\bibnamefont
  {Pezz{\`e}}}, \bibinfo {author} {\bibfnamefont {A.}~\bibnamefont {Smerzi}},
  \bibinfo {author} {\bibfnamefont {M.~K.}\ \bibnamefont {Oberthaler}},
  \bibinfo {author} {\bibfnamefont {R.}~\bibnamefont {Schmied}}, \ and\
  \bibinfo {author} {\bibfnamefont {P.}~\bibnamefont {Treutlein}},\ }\href
  {\doibase 10.1103/RevModPhys.90.035005} {\bibfield  {journal} {\bibinfo
  {journal} {Reviews of Modern Physics}\ }\textbf {\bibinfo {volume} {90}},\
  \bibinfo {pages} {035005} (\bibinfo {year} {2018})}\BibitemShut {NoStop}%
\bibitem [{\citenamefont {Degen}\ \emph {et~al.}(2017)\citenamefont {Degen},
  \citenamefont {Reinhard},\ and\ \citenamefont {Cappellaro}}]{Degen2017}%
  \BibitemOpen
  \bibfield  {author} {\bibinfo {author} {\bibfnamefont {C.~L.}\ \bibnamefont
  {Degen}}, \bibinfo {author} {\bibfnamefont {F.}~\bibnamefont {Reinhard}}, \
  and\ \bibinfo {author} {\bibfnamefont {P.}~\bibnamefont {Cappellaro}},\
  }\href {\doibase 10.1103/RevModPhys.89.035002} {\bibfield  {journal}
  {\bibinfo  {journal} {Reviews of Modern Physics}\ }\textbf {\bibinfo {volume}
  {89}},\ \bibinfo {pages} {035002} (\bibinfo {year} {2017})}\BibitemShut
  {NoStop}%
\bibitem [{\citenamefont {Szigeti}\ \emph {et~al.}(2021)\citenamefont
  {Szigeti}, \citenamefont {Hosten},\ and\ \citenamefont
  {Haine}}]{Szigeti2021}%
  \BibitemOpen
  \bibfield  {author} {\bibinfo {author} {\bibfnamefont {S.~S.}\ \bibnamefont
  {Szigeti}}, \bibinfo {author} {\bibfnamefont {O.}~\bibnamefont {Hosten}}, \
  and\ \bibinfo {author} {\bibfnamefont {S.~A.}\ \bibnamefont {Haine}},\ }\href
  {\doibase 10.1063/5.0050235} {\bibfield  {journal} {\bibinfo  {journal}
  {Applied Physics Letters}\ }\textbf {\bibinfo {volume} {118}},\ \bibinfo
  {pages} {140501} (\bibinfo {year} {2021})}\BibitemShut {NoStop}%
\bibitem [{\citenamefont {Giovannetti}\ \emph {et~al.}(2006)\citenamefont
  {Giovannetti}, \citenamefont {Lloyd},\ and\ \citenamefont
  {Maccone}}]{Giovannetti2006}%
  \BibitemOpen
  \bibfield  {author} {\bibinfo {author} {\bibfnamefont {V.}~\bibnamefont
  {Giovannetti}}, \bibinfo {author} {\bibfnamefont {S.}~\bibnamefont {Lloyd}},
  \ and\ \bibinfo {author} {\bibfnamefont {L.}~\bibnamefont {Maccone}},\
  }\href@noop {} {\bibfield  {journal} {\bibinfo  {journal} {Phys. Rev. Lett.}\
  }\textbf {\bibinfo {volume} {96}},\ \bibinfo {pages} {010401} (\bibinfo
  {year} {2006})}\BibitemShut {NoStop}%
\bibitem [{\citenamefont {Lee}(2006)}]{Lee2006}%
  \BibitemOpen
  \bibfield  {author} {\bibinfo {author} {\bibfnamefont {C.}~\bibnamefont
  {Lee}},\ }\href {\doibase 10.1103/PhysRevLett.97.150402} {\bibfield
  {journal} {\bibinfo  {journal} {Physical Review Letters}\ }\textbf {\bibinfo
  {volume} {97}},\ \bibinfo {pages} {150402} (\bibinfo {year}
  {2006})}\BibitemShut {NoStop}%
\bibitem [{\citenamefont {Pezz{\'e}}\ and\ \citenamefont
  {Smerzi}(2009)}]{Pezze2009}%
  \BibitemOpen
  \bibfield  {author} {\bibinfo {author} {\bibfnamefont {L.}~\bibnamefont
  {Pezz{\'e}}}\ and\ \bibinfo {author} {\bibfnamefont {A.}~\bibnamefont
  {Smerzi}},\ }\href {\doibase 10.1103/PhysRevLett.102.100401} {\bibfield
  {journal} {\bibinfo  {journal} {Physical Review Letters}\ }\textbf {\bibinfo
  {volume} {102}},\ \bibinfo {pages} {100401} (\bibinfo {year}
  {2009})}\BibitemShut {NoStop}%
\bibitem [{\citenamefont {Ockeloen}\ \emph {et~al.}(2013)\citenamefont
  {Ockeloen}, \citenamefont {Schmied}, \citenamefont {Riedel},\ and\
  \citenamefont {Treutlein}}]{Ockeloen2013}%
  \BibitemOpen
  \bibfield  {author} {\bibinfo {author} {\bibfnamefont {C.~F.}\ \bibnamefont
  {Ockeloen}}, \bibinfo {author} {\bibfnamefont {R.}~\bibnamefont {Schmied}},
  \bibinfo {author} {\bibfnamefont {M.~F.}\ \bibnamefont {Riedel}}, \ and\
  \bibinfo {author} {\bibfnamefont {P.}~\bibnamefont {Treutlein}},\ }\href
  {\doibase 10.1103/PhysRevLett.111.143001} {\bibfield  {journal} {\bibinfo
  {journal} {Physical Review Letters}\ }\textbf {\bibinfo {volume} {111}},\
  \bibinfo {pages} {143001} (\bibinfo {year} {2013})}\BibitemShut {NoStop}%
\bibitem [{\citenamefont {Hosten}\ \emph {et~al.}(2016)\citenamefont {Hosten},
  \citenamefont {Krishnakumar}, \citenamefont {Engelsen},\ and\ \citenamefont
  {Kasevich}}]{Hosten2016}%
  \BibitemOpen
  \bibfield  {author} {\bibinfo {author} {\bibfnamefont {O.}~\bibnamefont
  {Hosten}}, \bibinfo {author} {\bibfnamefont {R.}~\bibnamefont
  {Krishnakumar}}, \bibinfo {author} {\bibfnamefont {N.~J.}\ \bibnamefont
  {Engelsen}}, \ and\ \bibinfo {author} {\bibfnamefont {M.~A.}\ \bibnamefont
  {Kasevich}},\ }\href {\doibase 10.1126/science.aaf3397} {\bibfield  {journal}
  {\bibinfo  {journal} {Science}\ }\textbf {\bibinfo {volume} {352}},\ \bibinfo
  {pages} {1552} (\bibinfo {year} {2016})}\BibitemShut {NoStop}%
\bibitem [{\citenamefont {Szigeti}\ \emph {et~al.}(2020)\citenamefont
  {Szigeti}, \citenamefont {Nolan}, \citenamefont {Close},\ and\ \citenamefont
  {Haine}}]{Szigeti2020}%
  \BibitemOpen
  \bibfield  {author} {\bibinfo {author} {\bibfnamefont {S.~S.}\ \bibnamefont
  {Szigeti}}, \bibinfo {author} {\bibfnamefont {S.~P.}\ \bibnamefont {Nolan}},
  \bibinfo {author} {\bibfnamefont {J.~D.}\ \bibnamefont {Close}}, \ and\
  \bibinfo {author} {\bibfnamefont {S.~A.}\ \bibnamefont {Haine}},\ }\href
  {\doibase 10.1103/PhysRevLett.125.100402} {\bibfield  {journal} {\bibinfo
  {journal} {Physical Review Letters}\ }\textbf {\bibinfo {volume} {125}},\
  \bibinfo {pages} {100402} (\bibinfo {year} {2020})}\BibitemShut {NoStop}%
\bibitem [{\citenamefont {Giovannetti}(2004)}]{Giovannetti2004}%
  \BibitemOpen
  \bibfield  {author} {\bibinfo {author} {\bibfnamefont {V.}~\bibnamefont
  {Giovannetti}},\ }\href {\doibase 10.1126/science.1104149} {\bibfield
  {journal} {\bibinfo  {journal} {Science}\ }\textbf {\bibinfo {volume}
  {306}},\ \bibinfo {pages} {1330} (\bibinfo {year} {2004})}\BibitemShut
  {NoStop}%
\bibitem [{\citenamefont {Kitagawa}\ and\ \citenamefont
  {Ueda}(1993)}]{Kitagawa1993}%
  \BibitemOpen
  \bibfield  {author} {\bibinfo {author} {\bibfnamefont {M.}~\bibnamefont
  {Kitagawa}}\ and\ \bibinfo {author} {\bibfnamefont {M.}~\bibnamefont
  {Ueda}},\ }\href {\doibase 10.1103/PhysRevA.47.5138} {\bibfield  {journal}
  {\bibinfo  {journal} {Physical Review A}\ }\textbf {\bibinfo {volume} {47}},\
  \bibinfo {pages} {5138} (\bibinfo {year} {1993})}\BibitemShut {NoStop}%
\bibitem [{\citenamefont {Wineland}\ \emph {et~al.}(1992)\citenamefont
  {Wineland}, \citenamefont {Bollinger}, \citenamefont {Itano}, \citenamefont
  {Moore},\ and\ \citenamefont {Heinzen}}]{Wineland1992}%
  \BibitemOpen
  \bibfield  {author} {\bibinfo {author} {\bibfnamefont {D.~J.}\ \bibnamefont
  {Wineland}}, \bibinfo {author} {\bibfnamefont {J.~J.}\ \bibnamefont
  {Bollinger}}, \bibinfo {author} {\bibfnamefont {W.~M.}\ \bibnamefont
  {Itano}}, \bibinfo {author} {\bibfnamefont {F.~L.}\ \bibnamefont {Moore}}, \
  and\ \bibinfo {author} {\bibfnamefont {D.~J.}\ \bibnamefont {Heinzen}},\
  }\href {\doibase 10.1103/PhysRevA.46.R6797} {\bibfield  {journal} {\bibinfo
  {journal} {Physical Review A}\ }\textbf {\bibinfo {volume} {46}},\ \bibinfo
  {pages} {R6797} (\bibinfo {year} {1992})}\BibitemShut {NoStop}%
\bibitem [{\citenamefont {Wineland}\ \emph {et~al.}(1994)\citenamefont
  {Wineland}, \citenamefont {Bollinger}, \citenamefont {Itano},\ and\
  \citenamefont {Heinzen}}]{Wineland1994}%
  \BibitemOpen
  \bibfield  {author} {\bibinfo {author} {\bibfnamefont {D.~J.}\ \bibnamefont
  {Wineland}}, \bibinfo {author} {\bibfnamefont {J.~J.}\ \bibnamefont
  {Bollinger}}, \bibinfo {author} {\bibfnamefont {W.~M.}\ \bibnamefont
  {Itano}}, \ and\ \bibinfo {author} {\bibfnamefont {D.~J.}\ \bibnamefont
  {Heinzen}},\ }\href {\doibase 10.1103/PhysRevA.50.67} {\bibfield  {journal}
  {\bibinfo  {journal} {Physical Review A}\ }\textbf {\bibinfo {volume} {50}},\
  \bibinfo {pages} {67} (\bibinfo {year} {1994})}\BibitemShut {NoStop}%
\bibitem [{\citenamefont {Ma}\ \emph {et~al.}(2011)\citenamefont {Ma},
  \citenamefont {Wang}, \citenamefont {Sun},\ and\ \citenamefont
  {Nori}}]{Ma2011}%
  \BibitemOpen
  \bibfield  {author} {\bibinfo {author} {\bibfnamefont {J.}~\bibnamefont
  {Ma}}, \bibinfo {author} {\bibfnamefont {X.}~\bibnamefont {Wang}}, \bibinfo
  {author} {\bibfnamefont {C.}~\bibnamefont {Sun}}, \ and\ \bibinfo {author}
  {\bibfnamefont {F.}~\bibnamefont {Nori}},\ }\href {\doibase
  10.1016/j.physrep.2011.08.003} {\bibfield  {journal} {\bibinfo  {journal}
  {Physics Reports}\ }\textbf {\bibinfo {volume} {509}},\ \bibinfo {pages} {89}
  (\bibinfo {year} {2011})}\BibitemShut {NoStop}%
\bibitem [{\citenamefont {Sackett}(2010)}]{Sackett2010}%
  \BibitemOpen
  \bibfield  {author} {\bibinfo {author} {\bibfnamefont {C.~A.}\ \bibnamefont
  {Sackett}},\ }\href {\doibase 10.1038/4641133a} {\bibfield  {journal}
  {\bibinfo  {journal} {Nature}\ }\textbf {\bibinfo {volume} {464}},\ \bibinfo
  {pages} {1133} (\bibinfo {year} {2010})}\BibitemShut {NoStop}%
\bibitem [{\citenamefont {Monz}\ \emph {et~al.}(2011)\citenamefont {Monz},
  \citenamefont {Schindler}, \citenamefont {Barreiro}, \citenamefont {Chwalla},
  \citenamefont {Nigg}, \citenamefont {Coish}, \citenamefont {Harlander},
  \citenamefont {H{\"a}nsel}, \citenamefont {Hennrich},\ and\ \citenamefont
  {Blatt}}]{Monz2011}%
  \BibitemOpen
  \bibfield  {author} {\bibinfo {author} {\bibfnamefont {T.}~\bibnamefont
  {Monz}}, \bibinfo {author} {\bibfnamefont {P.}~\bibnamefont {Schindler}},
  \bibinfo {author} {\bibfnamefont {J.~T.}\ \bibnamefont {Barreiro}}, \bibinfo
  {author} {\bibfnamefont {M.}~\bibnamefont {Chwalla}}, \bibinfo {author}
  {\bibfnamefont {D.}~\bibnamefont {Nigg}}, \bibinfo {author} {\bibfnamefont
  {W.~A.}\ \bibnamefont {Coish}}, \bibinfo {author} {\bibfnamefont
  {M.}~\bibnamefont {Harlander}}, \bibinfo {author} {\bibfnamefont
  {W.}~\bibnamefont {H{\"a}nsel}}, \bibinfo {author} {\bibfnamefont
  {M.}~\bibnamefont {Hennrich}}, \ and\ \bibinfo {author} {\bibfnamefont
  {R.}~\bibnamefont {Blatt}},\ }\href {\doibase 10.1103/PhysRevLett.106.130506}
  {\bibfield  {journal} {\bibinfo  {journal} {Physical Review Letters}\
  }\textbf {\bibinfo {volume} {106}},\ \bibinfo {pages} {130506} (\bibinfo
  {year} {2011})}\BibitemShut {NoStop}%
\bibitem [{\citenamefont {Dunningham}\ \emph {et~al.}(2002)\citenamefont
  {Dunningham}, \citenamefont {Burnett},\ and\ \citenamefont
  {Barnett}}]{Dunningham2002}%
  \BibitemOpen
  \bibfield  {author} {\bibinfo {author} {\bibfnamefont {J.~A.}\ \bibnamefont
  {Dunningham}}, \bibinfo {author} {\bibfnamefont {K.}~\bibnamefont {Burnett}},
  \ and\ \bibinfo {author} {\bibfnamefont {S.~M.}\ \bibnamefont {Barnett}},\
  }\href {\doibase 10.1103/PhysRevLett.89.150401} {\bibfield  {journal}
  {\bibinfo  {journal} {Physical Review Letters}\ }\textbf {\bibinfo {volume}
  {89}},\ \bibinfo {pages} {150401} (\bibinfo {year} {2002})}\BibitemShut
  {NoStop}%
\bibitem [{\citenamefont {Campos}\ \emph {et~al.}(2003)\citenamefont {Campos},
  \citenamefont {Gerry},\ and\ \citenamefont {Benmoussa}}]{Campos2003}%
  \BibitemOpen
  \bibfield  {author} {\bibinfo {author} {\bibfnamefont {R.~A.}\ \bibnamefont
  {Campos}}, \bibinfo {author} {\bibfnamefont {C.~C.}\ \bibnamefont {Gerry}}, \
  and\ \bibinfo {author} {\bibfnamefont {A.}~\bibnamefont {Benmoussa}},\ }\href
  {\doibase 10.1103/PhysRevA.68.023810} {\bibfield  {journal} {\bibinfo
  {journal} {Physical Review A}\ }\textbf {\bibinfo {volume} {68}},\ \bibinfo
  {pages} {023810} (\bibinfo {year} {2003})}\BibitemShut {NoStop}%
\bibitem [{\citenamefont {Gerry}\ and\ \citenamefont
  {Mimih}(2010)}]{Gerry2010}%
  \BibitemOpen
  \bibfield  {author} {\bibinfo {author} {\bibfnamefont {C.~C.}\ \bibnamefont
  {Gerry}}\ and\ \bibinfo {author} {\bibfnamefont {J.}~\bibnamefont {Mimih}},\
  }\href {\doibase 10.1103/PhysRevA.82.013831} {\bibfield  {journal} {\bibinfo
  {journal} {Physical Review A}\ }\textbf {\bibinfo {volume} {82}},\ \bibinfo
  {pages} {013831} (\bibinfo {year} {2010})}\BibitemShut {NoStop}%
\bibitem [{\citenamefont {yu~Luo}\ \emph {et~al.}(2017)\citenamefont {yu~Luo},
  \citenamefont {quan Zou}, \citenamefont {na~Wu}, \citenamefont {Liu},
  \citenamefont {fei Han}, \citenamefont {Tey},\ and\ \citenamefont
  {Li}}]{Luo2017}%
  \BibitemOpen
  \bibfield  {author} {\bibinfo {author} {\bibfnamefont {X.}~\bibnamefont
  {yu~Luo}}, \bibinfo {author} {\bibfnamefont {Y.}~\bibnamefont {quan Zou}},
  \bibinfo {author} {\bibfnamefont {L.}~\bibnamefont {na~Wu}}, \bibinfo
  {author} {\bibfnamefont {Q.}~\bibnamefont {Liu}}, \bibinfo {author}
  {\bibfnamefont {M.}~\bibnamefont {fei Han}}, \bibinfo {author} {\bibfnamefont
  {M.~K.}\ \bibnamefont {Tey}}, \ and\ \bibinfo {author} {\bibfnamefont
  {Y.}~\bibnamefont {Li}},\ }\href {\doibase 10.1126/science.aag1106}
  {\bibfield  {journal} {\bibinfo  {journal} {Science}\ }\textbf {\bibinfo
  {volume} {355}},\ \bibinfo {pages} {620} (\bibinfo {year}
  {2017})}\BibitemShut {NoStop}%
\bibitem [{\citenamefont {Huang}\ \emph {et~al.}(2015)\citenamefont {Huang},
  \citenamefont {Qin}, \citenamefont {Zhong}, \citenamefont {Ke},\ and\
  \citenamefont {Lee}}]{Huang2015}%
  \BibitemOpen
  \bibfield  {author} {\bibinfo {author} {\bibfnamefont {J.}~\bibnamefont
  {Huang}}, \bibinfo {author} {\bibfnamefont {X.}~\bibnamefont {Qin}}, \bibinfo
  {author} {\bibfnamefont {H.}~\bibnamefont {Zhong}}, \bibinfo {author}
  {\bibfnamefont {Y.}~\bibnamefont {Ke}}, \ and\ \bibinfo {author}
  {\bibfnamefont {C.}~\bibnamefont {Lee}},\ }\href {\doibase 10.1038/srep17894}
  {\bibfield  {journal} {\bibinfo  {journal} {Scientific Reports}\ }\textbf
  {\bibinfo {volume} {5}},\ \bibinfo {pages} {17894} (\bibinfo {year}
  {2015})}\BibitemShut {NoStop}%
\bibitem [{\citenamefont {Lu}\ \emph {et~al.}(2019)\citenamefont {Lu},
  \citenamefont {Han}, \citenamefont {Zhuang}, \citenamefont {Ke},
  \citenamefont {Huang},\ and\ \citenamefont {Lee}}]{Lu2019}%
  \BibitemOpen
  \bibfield  {author} {\bibinfo {author} {\bibfnamefont {B.}~\bibnamefont
  {Lu}}, \bibinfo {author} {\bibfnamefont {C.-Y.}\ \bibnamefont {Han}},
  \bibinfo {author} {\bibfnamefont {M.}~\bibnamefont {Zhuang}}, \bibinfo
  {author} {\bibfnamefont {Y.-G.}\ \bibnamefont {Ke}}, \bibinfo {author}
  {\bibfnamefont {J.-H.}\ \bibnamefont {Huang}}, \ and\ \bibinfo {author}
  {\bibfnamefont {C.-H.}\ \bibnamefont {Lee}},\ }\href {\doibase
  10.7498/aps.68.20190147} {\bibfield  {journal} {\bibinfo  {journal} {Acta
  Physica Sinica}\ }\textbf {\bibinfo {volume} {68}},\ \bibinfo {pages}
  {040306} (\bibinfo {year} {2019})}\BibitemShut {NoStop}%
\bibitem [{\citenamefont {Huang}\ \emph {et~al.}(2014)\citenamefont {Huang},
  \citenamefont {Wu}, \citenamefont {Zhong},\ and\ \citenamefont
  {Lee}}]{Huang2014}%
  \BibitemOpen
  \bibfield  {author} {\bibinfo {author} {\bibfnamefont {J.}~\bibnamefont
  {Huang}}, \bibinfo {author} {\bibfnamefont {S.}~\bibnamefont {Wu}}, \bibinfo
  {author} {\bibfnamefont {H.}~\bibnamefont {Zhong}}, \ and\ \bibinfo {author}
  {\bibfnamefont {C.}~\bibnamefont {Lee}},\ }\href@noop {} {\emph {\bibinfo
  {title} {Quantum Metrology with Cold Atoms}}},\ Vol.~\bibinfo {volume} {2}\
  (\bibinfo {year} {2014})\ pp.\ \bibinfo {pages} {365--415}\BibitemShut
  {NoStop}%
\bibitem [{\citenamefont {Huang}\ \emph
  {et~al.}(2018{\natexlab{a}})\citenamefont {Huang}, \citenamefont {Zhuang},
  \citenamefont {Lu}, \citenamefont {Ke},\ and\ \citenamefont
  {Lee}}]{Huang2018-1}%
  \BibitemOpen
  \bibfield  {author} {\bibinfo {author} {\bibfnamefont {J.}~\bibnamefont
  {Huang}}, \bibinfo {author} {\bibfnamefont {M.}~\bibnamefont {Zhuang}},
  \bibinfo {author} {\bibfnamefont {B.}~\bibnamefont {Lu}}, \bibinfo {author}
  {\bibfnamefont {Y.}~\bibnamefont {Ke}}, \ and\ \bibinfo {author}
  {\bibfnamefont {C.}~\bibnamefont {Lee}},\ }\href {\doibase
  10.1103/PhysRevA.98.012129} {\bibfield  {journal} {\bibinfo  {journal}
  {Physical Review A}\ }\textbf {\bibinfo {volume} {98}},\ \bibinfo {pages}
  {012129} (\bibinfo {year} {2018}{\natexlab{a}})}\BibitemShut {NoStop}%
\bibitem [{\citenamefont {Nolan}\ \emph {et~al.}(2017)\citenamefont {Nolan},
  \citenamefont {Szigeti},\ and\ \citenamefont {Haine}}]{Nolan2017}%
  \BibitemOpen
  \bibfield  {author} {\bibinfo {author} {\bibfnamefont {S.~P.}\ \bibnamefont
  {Nolan}}, \bibinfo {author} {\bibfnamefont {S.~S.}\ \bibnamefont {Szigeti}},
  \ and\ \bibinfo {author} {\bibfnamefont {S.~A.}\ \bibnamefont {Haine}},\
  }\href {\doibase 10.1103/PhysRevLett.119.193601} {\bibfield  {journal}
  {\bibinfo  {journal} {Physical Review Letters}\ }\textbf {\bibinfo {volume}
  {119}},\ \bibinfo {pages} {1} (\bibinfo {year} {2017})}\BibitemShut {NoStop}%
\bibitem [{\citenamefont {Lee}(2009)}]{Lee2009}%
  \BibitemOpen
  \bibfield  {author} {\bibinfo {author} {\bibfnamefont {C.}~\bibnamefont
  {Lee}},\ }\href@noop {} {\bibfield  {journal} {\bibinfo  {journal} {Phys.
  Rev. Lett.}\ }\textbf {\bibinfo {volume} {102}},\ \bibinfo {pages} {070401}
  (\bibinfo {year} {2009})}\BibitemShut {NoStop}%
\bibitem [{\citenamefont {Gross}(2012)}]{Gross2012-1}%
  \BibitemOpen
  \bibfield  {author} {\bibinfo {author} {\bibfnamefont {C.}~\bibnamefont
  {Gross}},\ }\href {\doibase 10.1088/0953-4075/45/10/103001} {\bibfield
  {journal} {\bibinfo  {journal} {Journal of Physics B}\ }\textbf {\bibinfo
  {volume} {45}},\ \bibinfo {pages} {103001} (\bibinfo {year}
  {2012})}\BibitemShut {NoStop}%
\bibitem [{\citenamefont {Gross}(2010)}]{Gross2012-2}%
  \BibitemOpen
  \bibfield  {author} {\bibinfo {author} {\bibfnamefont {C.}~\bibnamefont
  {Gross}},\ }\href
  {http://books.google.de/books?id=lmtCqjtqg4EC&printsec=frontcover&dq=Spin+squeezing+and+non+linear+atom+interferometry+with+Bose+Einstein+condensates&cd=1&source=gbs_api}
  {\bibfield  {journal} {\bibinfo  {journal} {PhD dissertation}\ } (\bibinfo
  {year} {2010})}\BibitemShut {NoStop}%
\bibitem [{\citenamefont {Huang}\ \emph
  {et~al.}(2018{\natexlab{b}})\citenamefont {Huang}, \citenamefont {Zhuang},\
  and\ \citenamefont {Lee}}]{Huang2018-2}%
  \BibitemOpen
  \bibfield  {author} {\bibinfo {author} {\bibfnamefont {J.}~\bibnamefont
  {Huang}}, \bibinfo {author} {\bibfnamefont {M.}~\bibnamefont {Zhuang}}, \
  and\ \bibinfo {author} {\bibfnamefont {C.}~\bibnamefont {Lee}},\ }\href
  {\doibase 10.1103/PhysRevA.97.032116} {\bibfield  {journal} {\bibinfo
  {journal} {Physical Review A}\ }\textbf {\bibinfo {volume} {97}},\ \bibinfo
  {pages} {032116} (\bibinfo {year} {2018}{\natexlab{b}})}\BibitemShut
  {NoStop}%
\bibitem [{\citenamefont {Zhuang}\ \emph {et~al.}(2020)\citenamefont {Zhuang},
  \citenamefont {Huang}, \citenamefont {Ke},\ and\ \citenamefont
  {Lee}}]{Zhuang2020}%
  \BibitemOpen
  \bibfield  {author} {\bibinfo {author} {\bibfnamefont {M.}~\bibnamefont
  {Zhuang}}, \bibinfo {author} {\bibfnamefont {J.}~\bibnamefont {Huang}},
  \bibinfo {author} {\bibfnamefont {Y.}~\bibnamefont {Ke}}, \ and\ \bibinfo
  {author} {\bibfnamefont {C.}~\bibnamefont {Lee}},\ }\href {\doibase
  10.1002/andp.201900471} {\bibfield  {journal} {\bibinfo  {journal} {Annalen
  der Physik}\ }\textbf {\bibinfo {volume} {532}},\ \bibinfo {pages} {1900471}
  (\bibinfo {year} {2020})}\BibitemShut {NoStop}%
\bibitem [{\citenamefont {Trenkwalder}\ \emph {et~al.}(2016)\citenamefont
  {Trenkwalder}, \citenamefont {Spagnolli}, \citenamefont {Semeghini},
  \citenamefont {Coop}, \citenamefont {Landini}, \citenamefont {Castilho},
  \citenamefont {Pezz{\`e}}, \citenamefont {Modugno}, \citenamefont {Inguscio},
  \citenamefont {Smerzi},\ and\ \citenamefont {Fattori}}]{Trenkwalder2016}%
  \BibitemOpen
  \bibfield  {author} {\bibinfo {author} {\bibfnamefont {A.}~\bibnamefont
  {Trenkwalder}}, \bibinfo {author} {\bibfnamefont {G.}~\bibnamefont
  {Spagnolli}}, \bibinfo {author} {\bibfnamefont {G.}~\bibnamefont
  {Semeghini}}, \bibinfo {author} {\bibfnamefont {S.}~\bibnamefont {Coop}},
  \bibinfo {author} {\bibfnamefont {M.}~\bibnamefont {Landini}}, \bibinfo
  {author} {\bibfnamefont {P.}~\bibnamefont {Castilho}}, \bibinfo {author}
  {\bibfnamefont {L.}~\bibnamefont {Pezz{\`e}}}, \bibinfo {author}
  {\bibfnamefont {G.}~\bibnamefont {Modugno}}, \bibinfo {author} {\bibfnamefont
  {M.}~\bibnamefont {Inguscio}}, \bibinfo {author} {\bibfnamefont
  {A.}~\bibnamefont {Smerzi}}, \ and\ \bibinfo {author} {\bibfnamefont
  {M.}~\bibnamefont {Fattori}},\ }\href {\doibase 10.1038/nphys3743} {\bibfield
   {journal} {\bibinfo  {journal} {Nature Physics}\ }\textbf {\bibinfo {volume}
  {12}},\ \bibinfo {pages} {826} (\bibinfo {year} {2016})}\BibitemShut
  {NoStop}%
\bibitem [{\citenamefont {Jin}\ \emph {et~al.}(2009)\citenamefont {Jin},
  \citenamefont {Liu},\ and\ \citenamefont {Liu}}]{Jin2009}%
  \BibitemOpen
  \bibfield  {author} {\bibinfo {author} {\bibfnamefont {G.-R.}\ \bibnamefont
  {Jin}}, \bibinfo {author} {\bibfnamefont {Y.-C.}\ \bibnamefont {Liu}}, \ and\
  \bibinfo {author} {\bibfnamefont {W.-M.}\ \bibnamefont {Liu}},\ }\href
  {\doibase 10.1088/1367-2630/11/7/073049} {\bibfield  {journal} {\bibinfo
  {journal} {New Journal of Physics}\ }\textbf {\bibinfo {volume} {11}},\
  \bibinfo {pages} {073049} (\bibinfo {year} {2009})}\BibitemShut {NoStop}%
\bibitem [{\citenamefont {Strobel}\ \emph {et~al.}(2014)\citenamefont
  {Strobel}, \citenamefont {Muessel}, \citenamefont {Linnemann}, \citenamefont
  {Zibold}, \citenamefont {Hume}, \citenamefont {Pezze}, \citenamefont
  {Smerzi},\ and\ \citenamefont {Oberthaler}}]{Strobel2014}%
  \BibitemOpen
  \bibfield  {author} {\bibinfo {author} {\bibfnamefont {H.}~\bibnamefont
  {Strobel}}, \bibinfo {author} {\bibfnamefont {W.}~\bibnamefont {Muessel}},
  \bibinfo {author} {\bibfnamefont {D.}~\bibnamefont {Linnemann}}, \bibinfo
  {author} {\bibfnamefont {T.}~\bibnamefont {Zibold}}, \bibinfo {author}
  {\bibfnamefont {D.~B.}\ \bibnamefont {Hume}}, \bibinfo {author}
  {\bibfnamefont {L.}~\bibnamefont {Pezze}}, \bibinfo {author} {\bibfnamefont
  {A.}~\bibnamefont {Smerzi}}, \ and\ \bibinfo {author} {\bibfnamefont {M.~K.}\
  \bibnamefont {Oberthaler}},\ }\href {\doibase 10.1126/science.1250147}
  {\bibfield  {journal} {\bibinfo  {journal} {Science}\ }\textbf {\bibinfo
  {volume} {345}},\ \bibinfo {pages} {424} (\bibinfo {year}
  {2014})}\BibitemShut {NoStop}%
\bibitem [{\citenamefont {Davis}\ \emph {et~al.}(2016)\citenamefont {Davis},
  \citenamefont {Bentsen},\ and\ \citenamefont {Schleier-Smith}}]{Davis2016}%
  \BibitemOpen
  \bibfield  {author} {\bibinfo {author} {\bibfnamefont {E.}~\bibnamefont
  {Davis}}, \bibinfo {author} {\bibfnamefont {G.}~\bibnamefont {Bentsen}}, \
  and\ \bibinfo {author} {\bibfnamefont {M.}~\bibnamefont {Schleier-Smith}},\
  }\href {\doibase 10.1103/PhysRevLett.116.053601} {\bibfield  {journal}
  {\bibinfo  {journal} {Physical Review Letters}\ }\textbf {\bibinfo {volume}
  {116}},\ \bibinfo {pages} {053601} (\bibinfo {year} {2016})}\BibitemShut
  {NoStop}%
\bibitem [{\citenamefont {M\o{}lmer}\ and\ \citenamefont
  {S\o{}rensen}(1999)}]{Molmer1999}%
  \BibitemOpen
  \bibfield  {author} {\bibinfo {author} {\bibfnamefont {K.}~\bibnamefont
  {M\o{}lmer}}\ and\ \bibinfo {author} {\bibfnamefont {A.}~\bibnamefont
  {S\o{}rensen}},\ }\href {\doibase 10.1103/PhysRevLett.82.1835} {\bibfield
  {journal} {\bibinfo  {journal} {Phys. Rev. Lett.}\ }\textbf {\bibinfo
  {volume} {82}},\ \bibinfo {pages} {1835} (\bibinfo {year}
  {1999})}\BibitemShut {NoStop}%
\bibitem [{\citenamefont {Micheli}\ \emph {et~al.}(2003)\citenamefont
  {Micheli}, \citenamefont {Jaksch}, \citenamefont {Cirac},\ and\ \citenamefont
  {Zoller}}]{Micheli2003}%
  \BibitemOpen
  \bibfield  {author} {\bibinfo {author} {\bibfnamefont {A.}~\bibnamefont
  {Micheli}}, \bibinfo {author} {\bibfnamefont {D.}~\bibnamefont {Jaksch}},
  \bibinfo {author} {\bibfnamefont {J.~I.}\ \bibnamefont {Cirac}}, \ and\
  \bibinfo {author} {\bibfnamefont {P.}~\bibnamefont {Zoller}},\ }\href
  {\doibase 10.1103/PhysRevA.67.013607} {\bibfield  {journal} {\bibinfo
  {journal} {Physical Review A}\ }\textbf {\bibinfo {volume} {67}},\ \bibinfo
  {pages} {013607} (\bibinfo {year} {2003})}\BibitemShut {NoStop}%
\bibitem [{\citenamefont {Muessel}\ \emph {et~al.}(2015)\citenamefont
  {Muessel}, \citenamefont {Strobel}, \citenamefont {Linnemann}, \citenamefont
  {Zibold}, \citenamefont {Juli{\'a}-D{\'\i}az},\ and\ \citenamefont
  {Oberthaler}}]{Muessel2015}%
  \BibitemOpen
  \bibfield  {author} {\bibinfo {author} {\bibfnamefont {W.}~\bibnamefont
  {Muessel}}, \bibinfo {author} {\bibfnamefont {H.}~\bibnamefont {Strobel}},
  \bibinfo {author} {\bibfnamefont {D.}~\bibnamefont {Linnemann}}, \bibinfo
  {author} {\bibfnamefont {T.}~\bibnamefont {Zibold}}, \bibinfo {author}
  {\bibfnamefont {B.}~\bibnamefont {Juli{\'a}-D{\'\i}az}}, \ and\ \bibinfo
  {author} {\bibfnamefont {M.~K.}\ \bibnamefont {Oberthaler}},\ }\href
  {\doibase 10.1103/PhysRevA.92.023603} {\bibfield  {journal} {\bibinfo
  {journal} {Physical Review A - Atomic, Molecular, and Optical Physics}\
  }\textbf {\bibinfo {volume} {92}},\ \bibinfo {pages} {1} (\bibinfo {year}
  {2015})}\BibitemShut {NoStop}%
\bibitem [{\citenamefont {Mirkhalaf}\ \emph {et~al.}(2018)\citenamefont
  {Mirkhalaf}, \citenamefont {Nolan},\ and\ \citenamefont
  {Haine}}]{Mirkhalaf2018}%
  \BibitemOpen
  \bibfield  {author} {\bibinfo {author} {\bibfnamefont {S.~S.}\ \bibnamefont
  {Mirkhalaf}}, \bibinfo {author} {\bibfnamefont {S.~P.}\ \bibnamefont
  {Nolan}}, \ and\ \bibinfo {author} {\bibfnamefont {S.~A.}\ \bibnamefont
  {Haine}},\ }\href {\doibase 10.1103/PhysRevA.97.053618} {\bibfield  {journal}
  {\bibinfo  {journal} {Physical Review A}\ }\textbf {\bibinfo {volume} {97}},\
  \bibinfo {pages} {1} (\bibinfo {year} {2018})}\BibitemShut {NoStop}%
\bibitem [{\citenamefont {Sorelli}\ \emph {et~al.}(2019)\citenamefont
  {Sorelli}, \citenamefont {Gessner}, \citenamefont {Smerzi},\ and\
  \citenamefont {Pezz{\`e}}}]{Sorelli2019}%
  \BibitemOpen
  \bibfield  {author} {\bibinfo {author} {\bibfnamefont {G.}~\bibnamefont
  {Sorelli}}, \bibinfo {author} {\bibfnamefont {M.}~\bibnamefont {Gessner}},
  \bibinfo {author} {\bibfnamefont {A.}~\bibnamefont {Smerzi}}, \ and\ \bibinfo
  {author} {\bibfnamefont {L.}~\bibnamefont {Pezz{\`e}}},\ }\href {\doibase
  10.1103/PhysRevA.99.022329} {\bibfield  {journal} {\bibinfo  {journal}
  {Physical Review A}\ }\textbf {\bibinfo {volume} {99}},\ \bibinfo {pages} {1}
  (\bibinfo {year} {2019})}\BibitemShut {NoStop}%
\bibitem [{\citenamefont {Haine}\ and\ \citenamefont {Hope}(2020)}]{Haine2020}%
  \BibitemOpen
  \bibfield  {author} {\bibinfo {author} {\bibfnamefont {S.~A.}\ \bibnamefont
  {Haine}}\ and\ \bibinfo {author} {\bibfnamefont {J.~J.}\ \bibnamefont
  {Hope}},\ }\href {\doibase 10.1103/PhysRevLett.124.060402} {\bibfield
  {journal} {\bibinfo  {journal} {Physical Review Letters}\ }\textbf {\bibinfo
  {volume} {124}},\ \bibinfo {pages} {060402} (\bibinfo {year}
  {2020})}\BibitemShut {NoStop}%
\bibitem [{\citenamefont {Huo}\ \emph {et~al.}(2022)\citenamefont {Huo},
  \citenamefont {Min}, \citenamefont {Huang},\ and\ \citenamefont
  {Lee}}]{Huo2022}%
  \BibitemOpen
  \bibfield  {author} {\bibinfo {author} {\bibfnamefont {H.}~\bibnamefont
  {Huo}}, \bibinfo {author} {\bibfnamefont {Z.}~\bibnamefont {Min}}, \bibinfo
  {author} {\bibfnamefont {J.}~\bibnamefont {Huang}}, \ and\ \bibinfo {author}
  {\bibfnamefont {C.}~\bibnamefont {Lee}},\ }\href
  {http://iopscience.iop.org/article/10.1088/2058-9565/ac51af} {\bibfield
  {journal} {\bibinfo  {journal} {Quantum Science and Technology}\ } (\bibinfo
  {year} {2022})}\BibitemShut {NoStop}%
\bibitem [{\citenamefont {Omran}\ \emph {et~al.}(2019)\citenamefont {Omran},
  \citenamefont {Levine}, \citenamefont {Keesling}, \citenamefont {Semeghini},
  \citenamefont {Wang}, \citenamefont {Ebadi}, \citenamefont {Bernien},
  \citenamefont {Zibrov}, \citenamefont {Pichler}, \citenamefont {Choi},
  \citenamefont {Cui}, \citenamefont {Rossignolo}, \citenamefont {Rembold},
  \citenamefont {Montangero}, \citenamefont {Calarco}, \citenamefont {Endres},
  \citenamefont {Greiner}, \citenamefont {Vuleti{\'c}},\ and\ \citenamefont
  {Lukin}}]{Omran2019}%
  \BibitemOpen
  \bibfield  {author} {\bibinfo {author} {\bibfnamefont {A.}~\bibnamefont
  {Omran}}, \bibinfo {author} {\bibfnamefont {H.}~\bibnamefont {Levine}},
  \bibinfo {author} {\bibfnamefont {A.}~\bibnamefont {Keesling}}, \bibinfo
  {author} {\bibfnamefont {G.}~\bibnamefont {Semeghini}}, \bibinfo {author}
  {\bibfnamefont {T.~T.}\ \bibnamefont {Wang}}, \bibinfo {author}
  {\bibfnamefont {S.}~\bibnamefont {Ebadi}}, \bibinfo {author} {\bibfnamefont
  {H.}~\bibnamefont {Bernien}}, \bibinfo {author} {\bibfnamefont {A.~S.}\
  \bibnamefont {Zibrov}}, \bibinfo {author} {\bibfnamefont {H.}~\bibnamefont
  {Pichler}}, \bibinfo {author} {\bibfnamefont {S.}~\bibnamefont {Choi}},
  \bibinfo {author} {\bibfnamefont {J.}~\bibnamefont {Cui}}, \bibinfo {author}
  {\bibfnamefont {M.}~\bibnamefont {Rossignolo}}, \bibinfo {author}
  {\bibfnamefont {P.}~\bibnamefont {Rembold}}, \bibinfo {author} {\bibfnamefont
  {S.}~\bibnamefont {Montangero}}, \bibinfo {author} {\bibfnamefont
  {T.}~\bibnamefont {Calarco}}, \bibinfo {author} {\bibfnamefont
  {M.}~\bibnamefont {Endres}}, \bibinfo {author} {\bibfnamefont
  {M.}~\bibnamefont {Greiner}}, \bibinfo {author} {\bibfnamefont
  {V.}~\bibnamefont {Vuleti{\'c}}}, \ and\ \bibinfo {author} {\bibfnamefont
  {M.~D.}\ \bibnamefont {Lukin}},\ }\href {\doibase 10.1126/science.aax9743}
  {\bibfield  {journal} {\bibinfo  {journal} {Science}\ }\textbf {\bibinfo
  {volume} {365}},\ \bibinfo {pages} {570} (\bibinfo {year}
  {2019})}\BibitemShut {NoStop}%
\bibitem [{\citenamefont {Liu}\ \emph {et~al.}(2021)\citenamefont {Liu},
  \citenamefont {Zhang}, \citenamefont {Chen}, \citenamefont {Wang},\ and\
  \citenamefont {Yuan}}]{Liu2021}%
  \BibitemOpen
  \bibfield  {author} {\bibinfo {author} {\bibfnamefont {J.}~\bibnamefont
  {Liu}}, \bibinfo {author} {\bibfnamefont {M.}~\bibnamefont {Zhang}}, \bibinfo
  {author} {\bibfnamefont {H.}~\bibnamefont {Chen}}, \bibinfo {author}
  {\bibfnamefont {L.}~\bibnamefont {Wang}}, \ and\ \bibinfo {author}
  {\bibfnamefont {H.}~\bibnamefont {Yuan}},\ }\href {\doibase
  10.1002/qute.202100080} {\bibfield  {journal} {\bibinfo  {journal} {Advanced
  Quantum Technologies}\ ,\ \bibinfo {pages} {2100080}} (\bibinfo {year}
  {2021})}\BibitemShut {NoStop}%
\bibitem [{\citenamefont {Lin}\ \emph {et~al.}(2021)\citenamefont {Lin},
  \citenamefont {Ma},\ and\ \citenamefont {Sels}}]{Lin2021}%
  \BibitemOpen
  \bibfield  {author} {\bibinfo {author} {\bibfnamefont {C.}~\bibnamefont
  {Lin}}, \bibinfo {author} {\bibfnamefont {Y.}~\bibnamefont {Ma}}, \ and\
  \bibinfo {author} {\bibfnamefont {D.}~\bibnamefont {Sels}},\ }\href {\doibase
  10.1103/PhysRevA.103.052607} {\bibfield  {journal} {\bibinfo  {journal}
  {Phys. Rev. A}\ }\textbf {\bibinfo {volume} {103}},\ \bibinfo {pages}
  {052607} (\bibinfo {year} {2021})}\BibitemShut {NoStop}%
\bibitem [{\citenamefont {Kudra}\ \emph {et~al.}(2021)\citenamefont {Kudra},
  \citenamefont {Kervinen}, \citenamefont {Strandberg}, \citenamefont {Ahmed},
  \citenamefont {Scigliuzzo}, \citenamefont {Osman}, \citenamefont {Lozano},
  \citenamefont {Ferrini}, \citenamefont {Bylander}, \citenamefont {Kockum},
  \citenamefont {Quijandr{\'\i}a}, \citenamefont {Delsing},\ and\ \citenamefont
  {Gasparinetti}}]{Kudra2021}%
  \BibitemOpen
  \bibfield  {author} {\bibinfo {author} {\bibfnamefont {M.}~\bibnamefont
  {Kudra}}, \bibinfo {author} {\bibfnamefont {M.}~\bibnamefont {Kervinen}},
  \bibinfo {author} {\bibfnamefont {I.}~\bibnamefont {Strandberg}}, \bibinfo
  {author} {\bibfnamefont {S.}~\bibnamefont {Ahmed}}, \bibinfo {author}
  {\bibfnamefont {M.}~\bibnamefont {Scigliuzzo}}, \bibinfo {author}
  {\bibfnamefont {A.}~\bibnamefont {Osman}}, \bibinfo {author} {\bibfnamefont
  {D.~P.}\ \bibnamefont {Lozano}}, \bibinfo {author} {\bibfnamefont
  {G.}~\bibnamefont {Ferrini}}, \bibinfo {author} {\bibfnamefont
  {J.}~\bibnamefont {Bylander}}, \bibinfo {author} {\bibfnamefont {A.~F.}\
  \bibnamefont {Kockum}}, \bibinfo {author} {\bibfnamefont {F.}~\bibnamefont
  {Quijandr{\'\i}a}}, \bibinfo {author} {\bibfnamefont {P.}~\bibnamefont
  {Delsing}}, \ and\ \bibinfo {author} {\bibfnamefont {S.}~\bibnamefont
  {Gasparinetti}},\ }\href {http://arxiv.org/abs/2111.07965} {\ ,\ \bibinfo
  {pages} {1} (\bibinfo {year} {2021})}\BibitemShut {NoStop}%
\bibitem [{\citenamefont {Carrasco}\ \emph {et~al.}(2022)\citenamefont
  {Carrasco}, \citenamefont {Goerz}, \citenamefont {Li}, \citenamefont
  {Colombo}, \citenamefont {Vuleti{\'c}},\ and\ \citenamefont
  {Malinovsky}}]{Carrasco2022}%
  \BibitemOpen
  \bibfield  {author} {\bibinfo {author} {\bibfnamefont {S.~C.}\ \bibnamefont
  {Carrasco}}, \bibinfo {author} {\bibfnamefont {M.~H.}\ \bibnamefont {Goerz}},
  \bibinfo {author} {\bibfnamefont {Z.}~\bibnamefont {Li}}, \bibinfo {author}
  {\bibfnamefont {S.}~\bibnamefont {Colombo}}, \bibinfo {author} {\bibfnamefont
  {V.}~\bibnamefont {Vuleti{\'c}}}, \ and\ \bibinfo {author} {\bibfnamefont
  {V.~S.}\ \bibnamefont {Malinovsky}},\ }\href
  {http://arxiv.org/abs/2201.01744} {\  (\bibinfo {year} {2022})}\BibitemShut
  {NoStop}%
\bibitem [{\citenamefont {Riedel}\ \emph {et~al.}(2010)\citenamefont {Riedel},
  \citenamefont {B{\"o}hi}, \citenamefont {Li}, \citenamefont {Signnsch},
  \citenamefont {Sinatra},\ and\ \citenamefont {Treutlein}}]{Riedel2010}%
  \BibitemOpen
  \bibfield  {author} {\bibinfo {author} {\bibfnamefont {M.~F.}\ \bibnamefont
  {Riedel}}, \bibinfo {author} {\bibfnamefont {P.}~\bibnamefont {B{\"o}hi}},
  \bibinfo {author} {\bibfnamefont {Y.}~\bibnamefont {Li}}, \bibinfo {author}
  {\bibfnamefont {T.~W.~H.}\ \bibnamefont {Signnsch}}, \bibinfo {author}
  {\bibfnamefont {A.}~\bibnamefont {Sinatra}}, \ and\ \bibinfo {author}
  {\bibfnamefont {P.}~\bibnamefont {Treutlein}},\ }\href {\doibase
  10.1038/nature08988} {\bibfield  {journal} {\bibinfo  {journal} {Nature}\
  }\textbf {\bibinfo {volume} {464}},\ \bibinfo {pages} {1170} (\bibinfo {year}
  {2010})}\BibitemShut {NoStop}%
\bibitem [{\citenamefont {Gross}\ \emph {et~al.}(2010)\citenamefont {Gross},
  \citenamefont {Zibold}, \citenamefont {Nicklas}, \citenamefont {Est{\`e}ve},\
  and\ \citenamefont {Oberthaler}}]{Gross2010}%
  \BibitemOpen
  \bibfield  {author} {\bibinfo {author} {\bibfnamefont {C.}~\bibnamefont
  {Gross}}, \bibinfo {author} {\bibfnamefont {T.}~\bibnamefont {Zibold}},
  \bibinfo {author} {\bibfnamefont {E.}~\bibnamefont {Nicklas}}, \bibinfo
  {author} {\bibfnamefont {J.}~\bibnamefont {Est{\`e}ve}}, \ and\ \bibinfo
  {author} {\bibfnamefont {M.~K.}\ \bibnamefont {Oberthaler}},\ }\href
  {\doibase 10.1038/nature08919} {\bibfield  {journal} {\bibinfo  {journal}
  {Nature}\ }\textbf {\bibinfo {volume} {464}},\ \bibinfo {pages} {1165}
  (\bibinfo {year} {2010})}\BibitemShut {NoStop}%
\bibitem [{\citenamefont {Colombo}\ \emph {et~al.}(2021)\citenamefont
  {Colombo}, \citenamefont {Pedrozo-Peafiel}, \citenamefont {Adiyatullin},
  \citenamefont {Li}, \citenamefont {Mendez}, \citenamefont {Shu},\ and\
  \citenamefont {Vuletic}}]{Colombo2021}%
  \BibitemOpen
  \bibfield  {author} {\bibinfo {author} {\bibfnamefont {S.}~\bibnamefont
  {Colombo}}, \bibinfo {author} {\bibfnamefont {E.}~\bibnamefont
  {Pedrozo-Peafiel}}, \bibinfo {author} {\bibfnamefont {A.~F.}\ \bibnamefont
  {Adiyatullin}}, \bibinfo {author} {\bibfnamefont {Z.}~\bibnamefont {Li}},
  \bibinfo {author} {\bibfnamefont {E.}~\bibnamefont {Mendez}}, \bibinfo
  {author} {\bibfnamefont {C.}~\bibnamefont {Shu}}, \ and\ \bibinfo {author}
  {\bibfnamefont {V.}~\bibnamefont {Vuletic}},\ }\href@noop {} {\  (\bibinfo
  {year} {2021})},\ \Eprint {http://arxiv.org/abs/2106.03754} {arXiv:2106.03754
  [quant-ph]} \BibitemShut {NoStop}%
\bibitem [{\citenamefont {Li}\ \emph {et~al.}(2021)\citenamefont {Li},
  \citenamefont {Braverman}, \citenamefont {Colombo}, \citenamefont {Shu},
  \citenamefont {Kawasaki}, \citenamefont {Adiyatullin}, \citenamefont
  {Pedrozo-Pe{\~n}afiel}, \citenamefont {Mendez},\ and\ \citenamefont
  {Vuleti{\'c}}}]{Li2021}%
  \BibitemOpen
  \bibfield  {author} {\bibinfo {author} {\bibfnamefont {Z.}~\bibnamefont
  {Li}}, \bibinfo {author} {\bibfnamefont {B.}~\bibnamefont {Braverman}},
  \bibinfo {author} {\bibfnamefont {S.}~\bibnamefont {Colombo}}, \bibinfo
  {author} {\bibfnamefont {C.}~\bibnamefont {Shu}}, \bibinfo {author}
  {\bibfnamefont {A.}~\bibnamefont {Kawasaki}}, \bibinfo {author}
  {\bibfnamefont {A.}~\bibnamefont {Adiyatullin}}, \bibinfo {author}
  {\bibfnamefont {E.}~\bibnamefont {Pedrozo-Pe{\~n}afiel}}, \bibinfo {author}
  {\bibfnamefont {E.}~\bibnamefont {Mendez}}, \ and\ \bibinfo {author}
  {\bibfnamefont {V.}~\bibnamefont {Vuleti{\'c}}},\ }\href
  {http://arxiv.org/abs/2106.13234} {\  (\bibinfo {year} {2021})},\ \Eprint
  {http://arxiv.org/abs/2106.13234} {arXiv:2106.13234 [quant-ph]} \BibitemShut
  {NoStop}%
\bibitem [{\citenamefont {Ferrini}\ \emph {et~al.}(2010)\citenamefont
  {Ferrini}, \citenamefont {Spehner}, \citenamefont {Minguzzi},\ and\
  \citenamefont {Hekking}}]{Ferrini2010}%
  \BibitemOpen
  \bibfield  {author} {\bibinfo {author} {\bibfnamefont {G.}~\bibnamefont
  {Ferrini}}, \bibinfo {author} {\bibfnamefont {D.}~\bibnamefont {Spehner}},
  \bibinfo {author} {\bibfnamefont {A.}~\bibnamefont {Minguzzi}}, \ and\
  \bibinfo {author} {\bibfnamefont {F.~W.}\ \bibnamefont {Hekking}},\ }\href
  {\doibase 10.1103/PhysRevA.82.033621} {\bibfield  {journal} {\bibinfo
  {journal} {Physical Review A - Atomic, Molecular, and Optical Physics}\
  }\textbf {\bibinfo {volume} {82}} (\bibinfo {year} {2010}),\
  10.1103/PhysRevA.82.033621}\BibitemShut {NoStop}%
\bibitem [{\citenamefont {Spehner}\ \emph {et~al.}(2014)\citenamefont
  {Spehner}, \citenamefont {Pawlowski}, \citenamefont {Ferrini},\ and\
  \citenamefont {Minguzzi}}]{Spehner2014}%
  \BibitemOpen
  \bibfield  {author} {\bibinfo {author} {\bibfnamefont {D.}~\bibnamefont
  {Spehner}}, \bibinfo {author} {\bibfnamefont {K.}~\bibnamefont {Pawlowski}},
  \bibinfo {author} {\bibfnamefont {G.}~\bibnamefont {Ferrini}}, \ and\
  \bibinfo {author} {\bibfnamefont {A.}~\bibnamefont {Minguzzi}},\ }\href
  {\doibase 10.1140/epjb/e2014-50066-8} {\bibfield  {journal} {\bibinfo
  {journal} {The European Physical Journal B}\ }\textbf {\bibinfo {volume}
  {87}},\ \bibinfo {pages} {157} (\bibinfo {year} {2014})}\BibitemShut
  {NoStop}%
\bibitem [{\citenamefont {Braunstein}\ and\ \citenamefont
  {Caves}(1994)}]{Braunstein1994}%
  \BibitemOpen
  \bibfield  {author} {\bibinfo {author} {\bibfnamefont {S.~L.}\ \bibnamefont
  {Braunstein}}\ and\ \bibinfo {author} {\bibfnamefont {C.~M.}\ \bibnamefont
  {Caves}},\ }\href {\doibase 10.1103/PhysRevLett.72.3439} {\bibfield
  {journal} {\bibinfo  {journal} {Physical Review Letters}\ }\textbf {\bibinfo
  {volume} {72}},\ \bibinfo {pages} {3439} (\bibinfo {year}
  {1994})}\BibitemShut {NoStop}%
\bibitem [{\citenamefont {Guo}\ \emph {et~al.}(2021)\citenamefont {Guo},
  \citenamefont {Chen}, \citenamefont {Liu}, \citenamefont {Xue}, \citenamefont
  {Chen}, \citenamefont {Cao}, \citenamefont {Mao}, \citenamefont {Tey},\ and\
  \citenamefont {You}}]{Guo2021}%
  \BibitemOpen
  \bibfield  {author} {\bibinfo {author} {\bibfnamefont {S.-F.}\ \bibnamefont
  {Guo}}, \bibinfo {author} {\bibfnamefont {F.}~\bibnamefont {Chen}}, \bibinfo
  {author} {\bibfnamefont {Q.}~\bibnamefont {Liu}}, \bibinfo {author}
  {\bibfnamefont {M.}~\bibnamefont {Xue}}, \bibinfo {author} {\bibfnamefont
  {J.-J.}\ \bibnamefont {Chen}}, \bibinfo {author} {\bibfnamefont {J.-H.}\
  \bibnamefont {Cao}}, \bibinfo {author} {\bibfnamefont {T.-W.}\ \bibnamefont
  {Mao}}, \bibinfo {author} {\bibfnamefont {M.~K.}\ \bibnamefont {Tey}}, \ and\
  \bibinfo {author} {\bibfnamefont {L.}~\bibnamefont {You}},\ }\href {\doibase
  10.1103/PhysRevLett.126.060401} {\bibfield  {journal} {\bibinfo  {journal}
  {Physical Review Letters}\ }\textbf {\bibinfo {volume} {126}},\ \bibinfo
  {pages} {060401} (\bibinfo {year} {2021})}\BibitemShut {NoStop}%
\bibitem [{\citenamefont {Zhou}\ \emph {et~al.}(2020)\citenamefont {Zhou},
  \citenamefont {Wang}, \citenamefont {Choi}, \citenamefont {Pichler},\ and\
  \citenamefont {Lukin}}]{Zhou2020}%
  \BibitemOpen
  \bibfield  {author} {\bibinfo {author} {\bibfnamefont {L.}~\bibnamefont
  {Zhou}}, \bibinfo {author} {\bibfnamefont {S.~T.}\ \bibnamefont {Wang}},
  \bibinfo {author} {\bibfnamefont {S.}~\bibnamefont {Choi}}, \bibinfo {author}
  {\bibfnamefont {H.}~\bibnamefont {Pichler}}, \ and\ \bibinfo {author}
  {\bibfnamefont {M.~D.}\ \bibnamefont {Lukin}},\ }\href {\doibase
  10.1103/PhysRevX.10.021067} {\bibfield  {journal} {\bibinfo  {journal}
  {Physical Review X}\ }\textbf {\bibinfo {volume} {10}},\ \bibinfo {pages}
  {21067} (\bibinfo {year} {2020})}\BibitemShut {NoStop}%
\bibitem [{\citenamefont {Titum}\ \emph {et~al.}(2021)\citenamefont {Titum},
  \citenamefont {Schultz}, \citenamefont {Seif}, \citenamefont {Quiroz},\ and\
  \citenamefont {Clader}}]{Titum2021}%
  \BibitemOpen
  \bibfield  {author} {\bibinfo {author} {\bibfnamefont {P.}~\bibnamefont
  {Titum}}, \bibinfo {author} {\bibfnamefont {K.}~\bibnamefont {Schultz}},
  \bibinfo {author} {\bibfnamefont {A.}~\bibnamefont {Seif}}, \bibinfo {author}
  {\bibfnamefont {G.}~\bibnamefont {Quiroz}}, \ and\ \bibinfo {author}
  {\bibfnamefont {B.~D.}\ \bibnamefont {Clader}},\ }\href {\doibase
  10.1038/s41534-021-00383-5} {\bibfield  {journal} {\bibinfo  {journal} {npj
  Quantum Information}\ }\textbf {\bibinfo {volume} {7}},\ \bibinfo {pages}
  {53} (\bibinfo {year} {2021})}\BibitemShut {NoStop}%
\bibitem [{\citenamefont {Kaubruegger}\ \emph {et~al.}(2021)\citenamefont
  {Kaubruegger}, \citenamefont {Vasilyev}, \citenamefont {Schulte},
  \citenamefont {Hammerer},\ and\ \citenamefont {Zoller}}]{Kaubruegger2021}%
  \BibitemOpen
  \bibfield  {author} {\bibinfo {author} {\bibfnamefont {R.}~\bibnamefont
  {Kaubruegger}}, \bibinfo {author} {\bibfnamefont {D.~V.}\ \bibnamefont
  {Vasilyev}}, \bibinfo {author} {\bibfnamefont {M.}~\bibnamefont {Schulte}},
  \bibinfo {author} {\bibfnamefont {K.}~\bibnamefont {Hammerer}}, \ and\
  \bibinfo {author} {\bibfnamefont {P.}~\bibnamefont {Zoller}},\ }\href
  {\doibase 10.1103/PhysRevX.11.041045} {\bibfield  {journal} {\bibinfo
  {journal} {Physical Review X}\ }\textbf {\bibinfo {volume} {11}},\ \bibinfo
  {pages} {041045} (\bibinfo {year} {2021})}\BibitemShut {NoStop}%
\end{thebibliography}

\end{document}